\documentclass[superscriptaddress,groupedaddress,amsmath,amssymb,aps,12pt,physrev]{revtex4-2}

\usepackage{graphicx}    
\usepackage{subfig}      
\usepackage{float}       

\usepackage{dcolumn,multirow}
\usepackage{bm}
\usepackage{amsmath,amssymb}
\usepackage{hyperref}
\usepackage{xcolor, soul}
\usepackage{placeins}

\begin{document}

\title{\textbf{Revisiting Phase Transitions of Yttrium: Insights from Density Functional Theory} 
}%
\author{Paras Patel$^{1}$, Madhavi H. Dalsaniya$^{2,3}$, Saurav Patel$^{1}$, Dominik Kurzydłowski$^{3}$, Krzysztof J. Kurzydłowski$^{2,4}$, Prafulla K. Jha$^{1,*}$}

\affiliation{$^1$Department of Physics, Faculty of Science, The Maharaja Sayajirao University of Baroda, Vadodara-390002, Gujarat, India.}
\affiliation{$^2$Faculty of Materials Science and Engineering, Warsaw University of Technology, Wołoska 141, 02-507, Warsaw, Poland.}
\affiliation{$^3$Faculty of Mathematics and Natural Sciences, Cardinal Stefan Wyszyński University in Warsaw, 01-038 Warsaw, Poland.}
\affiliation{$^4$Faculty of Mechanical Engineering, Bialystok University of Technology, Wiejska 45C, 15-351, Bialystok, Poland.}

\affiliation{$^*$Corresponding author: \textit{prafullaj@yahoo.com}}

\begin{abstract}
	\fontsize{10}{10}\selectfont
Understanding the mechanism of structural phase transitions in rare-earth elements is a fundamental challenge in condensed matter physics, with significant implications for materials science applications. In this study, we present a systematic investigation on the phase transitions of yttrium under low-pressure conditions ($<$30 GPa) focusing on the hcp, Sm-type, and dhcp phases. A comparative analysis of the generalized gradient approximation (GGA) and meta-GGA functionals reveals that the PBE-GGA functional significantly underestimates the phase transition pressures, whereas the r$^2$SCAN functional provides accurate predictions of phase transition pressures which are in excellent agreement with experimental data. The results confirm that the phase transitions in yttrium are driven by vibrational instabilities, as evidenced by the emergence of soft acoustic modes in the phonon dispersion curves for the hcp and Sm-type phase. Elastic properties calculations further confirm mechanical softening at the phase boundaries, particularly in the hcp phase, suggesting a strong correlation between elastic instability and structural transitions. These findings suggest that the emergence of soft modes in the phonon dispersion curves might be a key factor driving the structural phase transition in the rare earth materials.
	\normalsize
\end{abstract}

\maketitle

\section{\label{I}Introduction}

The remarkable structural and material properties of rare earth elements under extreme conditions have inspired extensive experimental and theoretical investigations over the years due to their pivotal role in high pressure science \cite{Benedict1986,Kruger1990,Grosshans1992,Lei2007,Soderlind2014}. The trivalent transition elements Yttrium (Y) and scandium (Sc), are grouped with the rare-earth elements due of their structural and electronic similarities with the lanthanide series, spanning from lanthanum (La) to lutetium (Lu) \cite{Holzapfel1995,McMahon2019}. Among these elements, yttrium stands out as a benchmark system, valued for its relatively simple electronic structure and absence of 4f electrons, making it an ideal candidate for exploring high-pressure behaviour and structural phase transitions \cite{Samudrala2012,Pace2020}. 

The trivalent lanthanides, with the exception of cerium (Ce), europium (Eu), and ytterbium (Yb), follow a consistent sequence of structural transitions under compression. These transitions involve variations in the stacking of close-packed atomic layers and follow the sequence: hexagonal close-packed ($hcp$) → samarium-type ($Sm$-$type$) → double hexagonal close-packed ($dhcp$) → face-centred cubic ($FCC$) → distorted face-centred cubic ($DFCC$) \cite{Johansson1975,Chesnut1998,Patterson2004,Samudrala2011}. Initially, these transitions were attributed to changes in $f-$band electron occupancy either due to increasing atomic number or applied pressure \cite{Johansson1975}. However, later studies revealed that yttrium, which has no $f$ electrons, undergoes an identical sequence of transitions that is quantitatively correlated with the pressure-induced transfer electrons from s to d states, driven by the d-band energy contribution to the total energy \cite{Duthie1977}.

Subsequently, yttrium serves as a reference system for understanding the mechanisms of the structural phase transitions of rare-earth metals \cite{Samudrala2012,Vohra1981,Grosshans1982}. At ambient conditions, yttrium crystallizes in an $hcp$ ($P6_3/mmc$) structure. Under compression, it transitions to a $Sm$-$type$ ($R\bar3 m$) phase at ~10 GPa, followed by a $dhcp$ ($P6_3/mmc$) phase at around 25 GPa \cite{Vohra1981} and subsequently to a $dfcc$ ($hR24$) at 50 GPa \cite{Grosshans1992,Samudrala2012}. At pressures near 100 GPa, yttrium undergoes another structural transformation, initially identified as leading to the monoclinic $C2/m$ phase \cite{Samudrala2012}. In 2011, Y. Chen et.al theoretically predicted the occurrence of $P6_222$ phase which is dynamically stable beyond 206 GPa \cite{Chen2011}. However, recent studies suggest that the phase observed above 100 GPa exhibits an orthorhombic $Fddd$ ($oF16$) structure, which is predicted to persist up to 524 GPa \cite{Buhot2020,Chen2012}. Yttrium is also notable for its superconducting properties. It exhibits a remarkably high superconducting transition temperature ($T_c$) of 19.5 K at 115 GPa \cite{Hamlin2007}, further highlighting its importance in high-pressure and superconductivity research.

Despite significant advancements, discrepancies persist between theoretical models and experimental observations, particularly in the lower-pressure range \cite{Li2019}. The predicted transition pressures for $hcp$-to-$Sm$-$type$ and $Sm$-$type$-to-$dhcp$ transitions often differ from experimental values, emphasizing the need for more accurate theoretical frameworks that can account for both electronic contributions and lattice dynamics. Additionally, while energetic considerations have been extensively studied, questions about the dynamic and thermodynamic stabilities of these phases remain unresolved. Density functional theory (DFT) has played a crucial role in understanding yttrium’s high-pressure behaviour \cite{Giannessi2024,Liu2023,Li2022}. Standard generalized gradient approximation ($GGA$) functionals, such as $Perdew-Burke-Ernzerhof$  ($PBE$), have provided valuable insights but often struggle to fully capture the subtleties of electronic correlations and lattice interactions \cite{Li2019,Li2022}. Recent advancements in exchange-correlation methodologies, such as $meta$-$GGA$ functionals ($SCAN$ and $r^2SCAN$), offer improved accuracy in predicting phase stability and transition pressures, providing a more reliable approach to studying these transitions \cite{Dalsaniya2022,Yang2020}. By leveraging these advanced computational tools, researchers are better equipped to bridge the gaps between theoretical and experimental findings.

Re-examining yttrium’s structural transitions under pressure is not only important for addressing these scientific challenges but also to understand their broader implications. Moreover, yttrium’s high-pressure phases are of interest for applications in   superconductivity \cite{Hamlin2007}, where understanding the interplay of structure, electronic properties, lattice dynamical properties, mechanical properties and phase stability under extreme conditions is critical. This study aims to address these challenges by integrating advanced computational approaches with comparison of previously reported experimental findings to develop a comprehensive understanding of yttrium’s phase transitions at lower pressure. By addressing discrepancies in transition pressures and evaluating the stability of various phases using advanced exchange-correlation functionals, this research aims to clarify the mechanisms governing structural evolution in yttrium and provide a framework for studying high-pressure behaviours in other rare-earth elements and materials.

\section{\label{II}METHODS}

The structural relaxations were performed using DFT within the $GGA$ ($PBE$) \cite{Perdew1996}, $meta$-$GGA$ ($SCAN$ \cite{Sun2015} and $r^2SCAN$ \cite{Furness2020}) approximations as implemented in the Vienna Ab initio simulation package (VASP-6.4.2) \cite{Kresse1996}. The projector augmented wave (PAW) pseudopotential \cite{Blochl1994} with $4s^2$, $4p^6$, $5s^2$ and $4d^1$ orbitals as valance configuration was chosen for ion-electron interaction. To consider the effect of Van der Waals interaction between the atoms of yttrium, DFT-D3 method with Becke-Johnson damping function \cite{Grimme2011} was applied in combination with GGA and meta-GGA functionals. A kinetic energy cut-off of 600 eV and uniform $\Gamma$-centred denser k-meshes with a resolution of $2\pi\times0.03$ \AA$^{-1}$ were used for precise energy calculations. The convergence criterion for energy and forces were set to 10$^{-8}$ eV and 10$^{-2}$ eV \AA$^{-1}$ respectively for electronic relaxation. The optimization of all structural parameters was carried out for a pressure range of 0 to 30 GPa within 2 GPa intervals. Phonon dispersion curves and elastic constant calculations were done using the finite difference method with a supercell approach as implemented in VASP. Mulliken charge population calculations were done using Lobster 5.0.0 \cite{Nelson2020}. While  structure were visualized using VESTA package \cite{Momma2011}.

\section{\label{III}RESULTS AND DISCUSSION}

\begin{figure}[h]
	\centering
	
	\begin{minipage}{0.45\columnwidth} 
		\raggedright 
		{\fontsize{12}{14}\selectfont \textbf{(a)}} \\  
		\includegraphics[width=\linewidth]{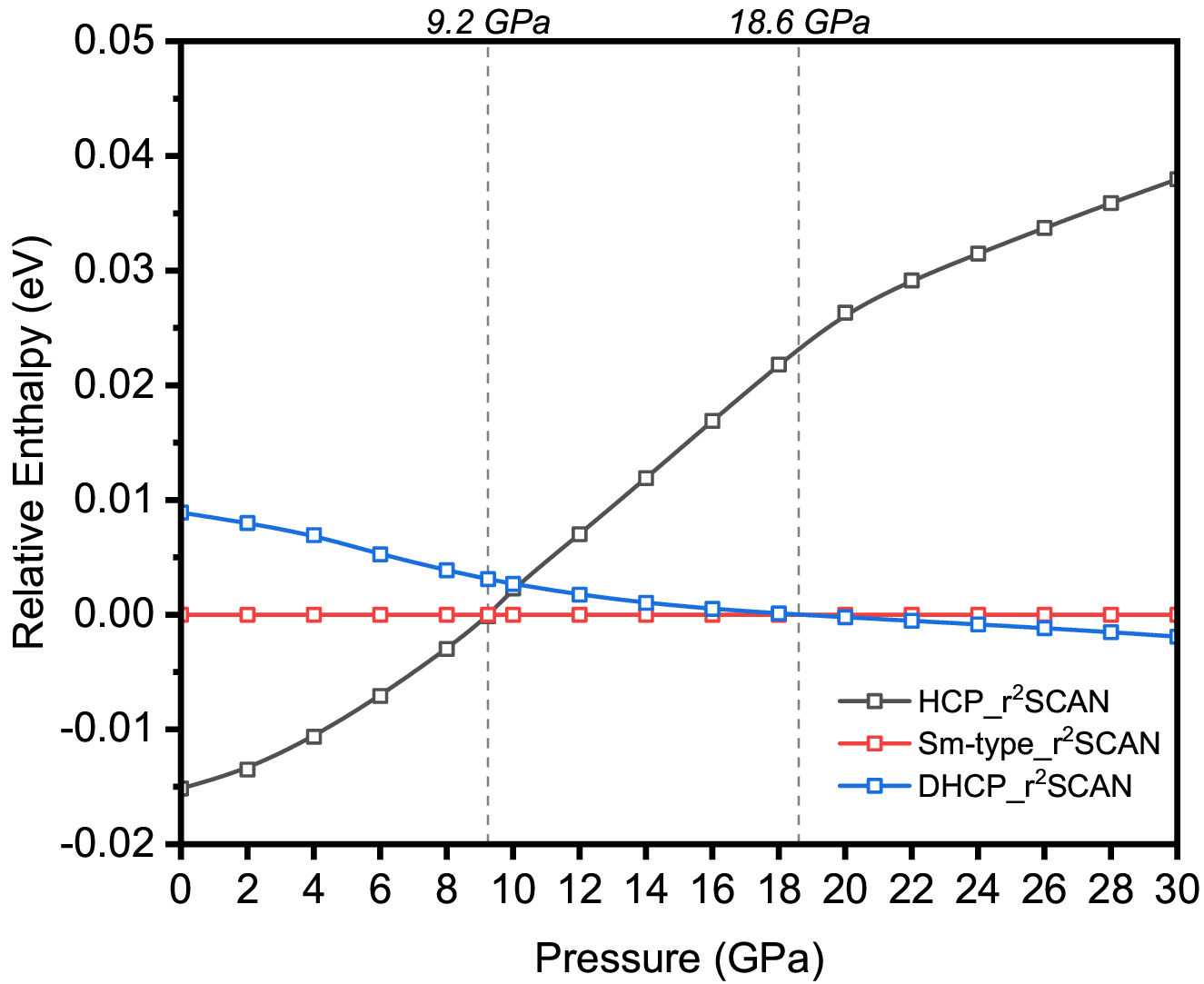} 
	\end{minipage}
	\hspace{2pt} 
	\begin{minipage}{0.45\columnwidth} 
		\raggedright 
		{\fontsize{12}{14}\selectfont \textbf{(b)}} \\  
		\includegraphics[width=\linewidth]{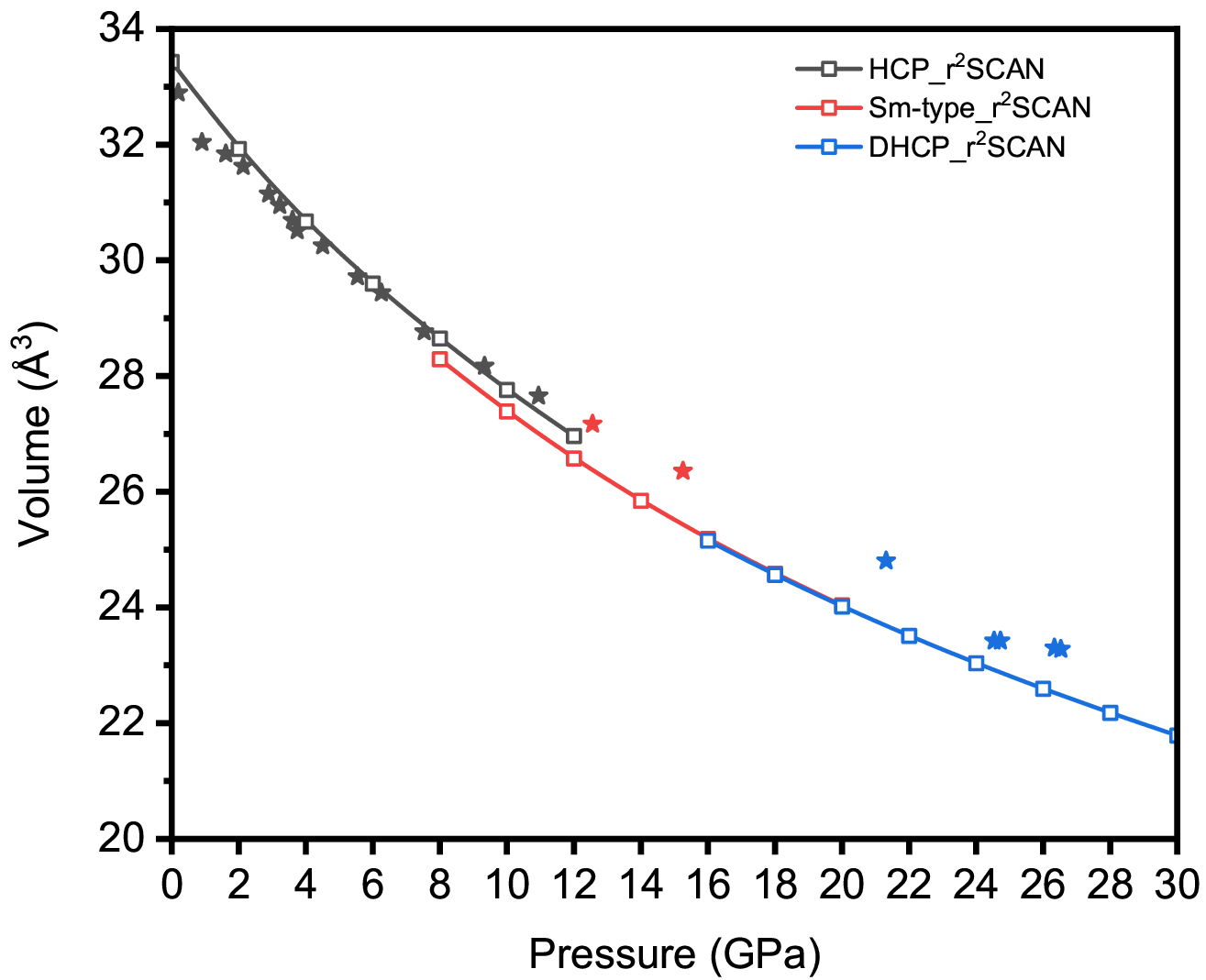} 
	\end{minipage}
	
	\caption{The pressure-enthalpy profile (a) and pressure-volume profile, compared with experimental values \cite{Pace2020} (represented by star symbols), (b) calculated using the $r^2SCAN$ functional.}
	
	\label{fig:1}
\end{figure}

\begin{figure}[b]
	\centering
	
	\begin{minipage}[b]{0.25\columnwidth}
		\raggedright
		{\fontsize{12}{14}\selectfont \textbf{(a)}} \\  
		\includegraphics[width=\linewidth]{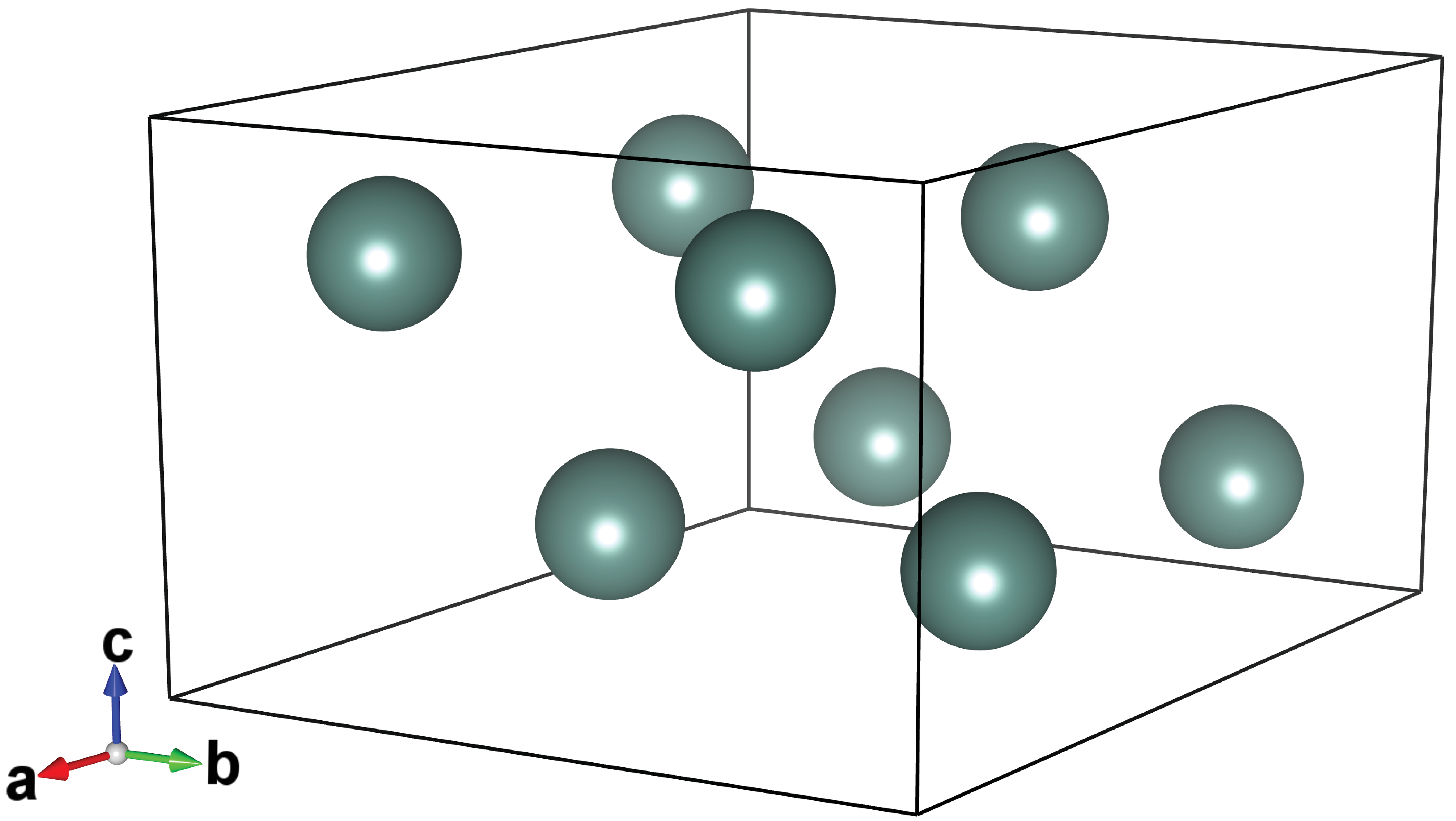}
	\end{minipage}
	\hspace{2pt} 
	\begin{minipage}[b]{0.25\columnwidth}
		\raggedright
		{\fontsize{12}{14}\selectfont \textbf{(b)}} \\  
		\includegraphics[width=\linewidth]{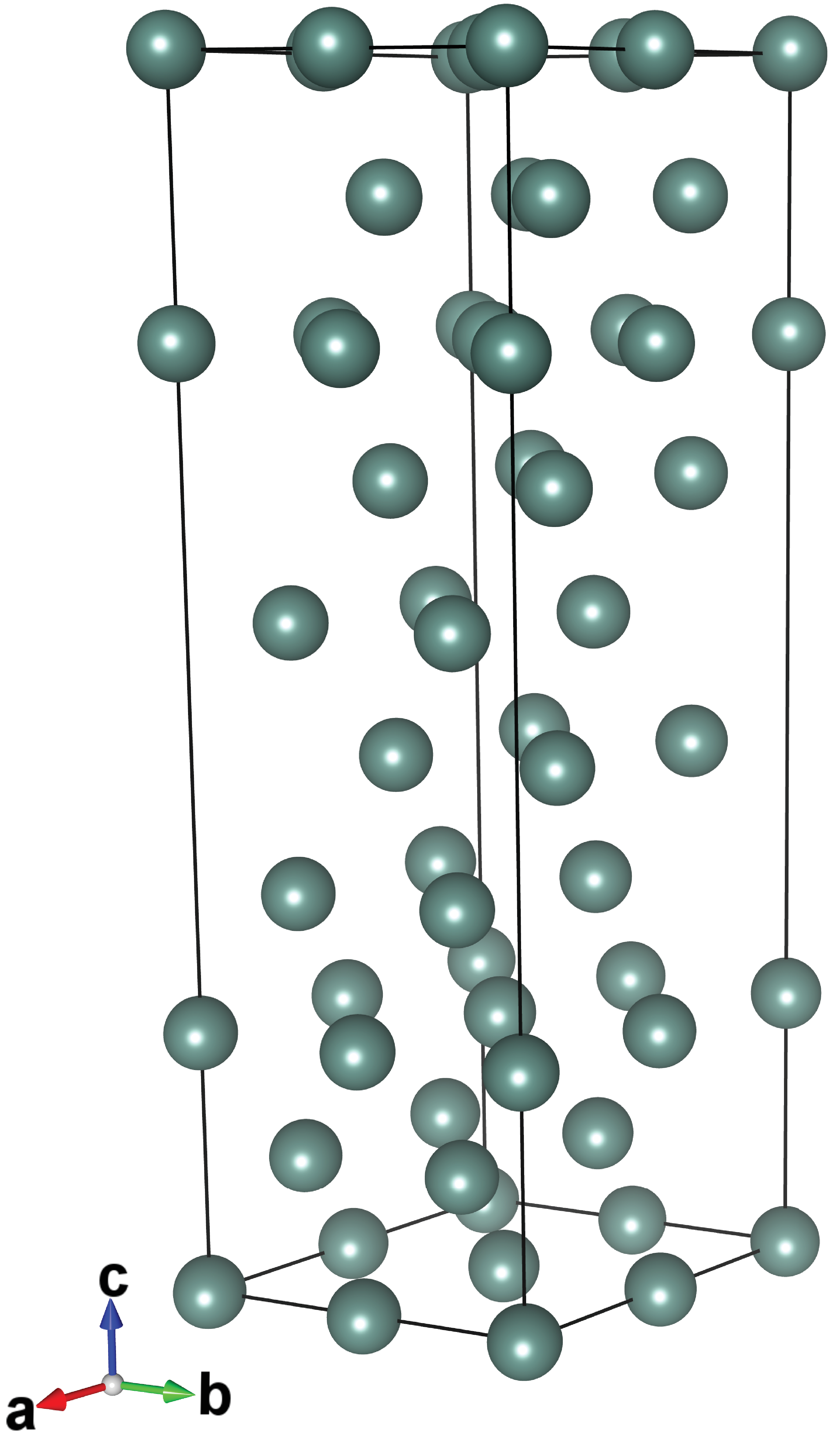}
	\end{minipage}
	\hspace{2pt} 
	\begin{minipage}[b]{0.25\columnwidth}
		\raggedright
		{\fontsize{12}{14}\selectfont \textbf{(c)}} \\  
		\includegraphics[width=\linewidth]{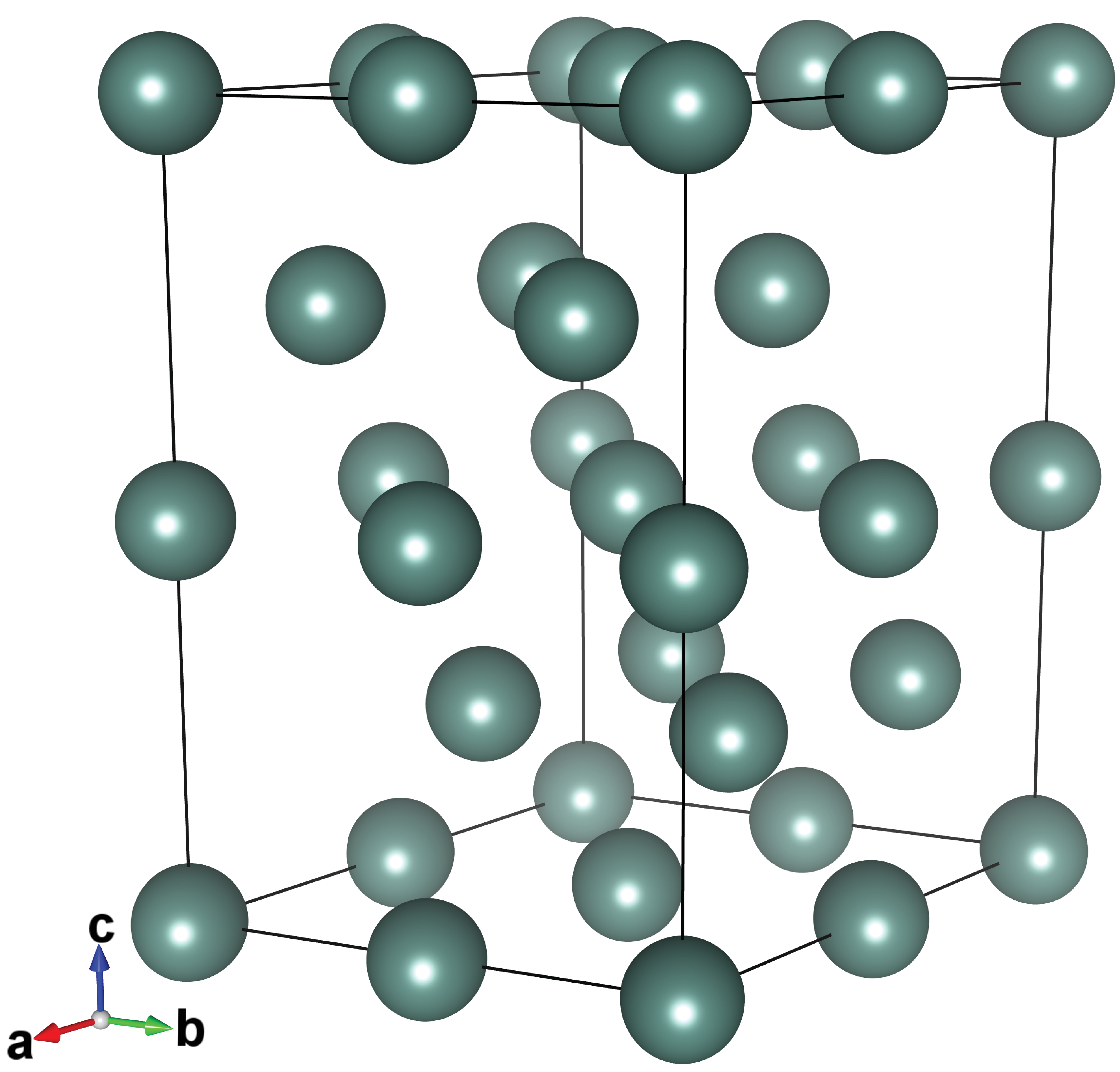}
	\end{minipage}
	
	\caption{Structural view (supercell) of hcp at 1 atm   (a), Sm-type at 10 GPa (b) and dhcp at 20 GPa (c) phase of yttrium.}
	\label{fig:2}
\end{figure}

Many structures have been proposed through both theoretical and experimental studies for yttrium. Since our focus is on the low pressure ($<$30 GPa) structural phase transitions, the selected candidate structures are $hcp$ ($P6_3/mmc-2$), $Sm$-$type$ ($R\bar3m$) and $dhcp$ ($P6_3/mmc-4$). Specifically, the first structural phase transition from $hcp$ to $Sm$-$type$ using GGA-PBE functional is observed at 0.5 GPa, while the second structural phase transition above 9 GPa. These structural phase transition sequences are consistent with the previously reported theoretical articles \cite{Chen2011,Li2019}. However, the predicted transition pressure significantly differs from the experimental results \cite{Samudrala2012}. Since the GGA functionals lack the inclusion of electronic-level kinetic energy contributions, which become increasingly significant under high-pressure conditions, they fail to accurately predict phase transitions, necessitating the use of more advance methods beyond GGA to correctly interpret the behaviour of yttrium. To address these challenges, we performed extensive computations by considering the meta-GGA functionals ($SCAN$ and $r^2SCAN$) along with inclusion of van der Waals (vdW) forces by considering DFT-D3 corrections. The calculations of $r^2SCAN$ provides good estimation for the transition pressure as evidenced from pressure-enthalpy profile presented in Fig. \ref{fig:1}a. The comparison of transition pressure based on enthalpy of structures is tabulated in Table \ref{table1}.
\begin{table}[t] 
	\caption{\label{table1}%
		Comparison of transition pressure (in GPa) using different methods.
	}
	\begin{ruledtabular}
		\fontsize{10}{10}\selectfont
		\begin{tabular}{lcc} 
			\textrm{Method} & 
			\textrm{$hcp \rightarrow Sm$-$type$} & 
			\textrm{$Sm$-$type \rightarrow dhcp$} \\
			\colrule
			Experimental\footnotemark[1] & $\sim$10 & $\sim$25 \\ 
			Experimental\footnotemark[2] & $\sim$13-16 & $\sim$25 \\ 
			GGA\footnotemark[3] & 0 & 7 \\ 
			LDA\footnotemark[3] & -4 & 3 \\
			PBE\footnotemark[4] & 0.5 & 8 \\
			Finnis-Sinclair & \multirow{2}{*}{10.5} & \multirow{2}{*}{13.5} \\
			interatomic potential\footnotemark[5] & & \\
			PBE-present & 2.8 & 7.5 \\
			PBE+D3-present & 0.9 & 5.2 \\
			\textbf{r$^2$SCAN-present} & \textbf{9.2} & \textbf{18.6} \\
			r$^2$SCAN+D3-present & 8 & 17.3 \\
			SCAN-present & 8 & 14 \\
		\end{tabular}
		\normalsize
	\end{ruledtabular}
	\footnotetext[1]{Reference \cite{Vohra1981}} 
	\footnotetext[2]{Reference \cite{Samudrala2012}}
	\footnotetext[3]{Reference \cite{Chen2011}}
	\footnotetext[4]{Reference \cite{Li2019}}
	\footnotetext[5]{Reference \cite{Liu2023}}
\end{table}
It is clear from the Table 1 that the primitive cell calculations do not exhibit any substantial improvement in the existing theoretical data. Furthermore, other functionals show numerical inaccuracies. Consequently, for refined ground-state energy calculations, supercells of size $2\times2\times1$ were employed, containing 8 atoms, 36 atoms, and 16 atoms for the $hcp$, $Sm$-$type$ and $dhcp$ phase, respectively (Fig. \ref{fig:2}).

The $PBE$ functional show improvement for the results for the first phase transition prediction which is at 2.8 GPa in comparison to the earlier reperformed value (0 GPa) however, predictions for the second phase transition remain inadequate (shown in Fig. S1a). In contrast, $r^2SCAN$ functional calculations reveal that the $hcp$ retains the lowest enthalpy up to 9.2 GPa, which is very close to the previously reported experimental value of  10 GPa \cite{Vohra1981}. B. Liu et al have predicted the first phase transition (at 10.5 GPa) close to the experimental transition by developing the Finnis-Sinclair interatomic potential through fitting of the experimental and first principles databases however, they failed in the second phase transition (13.5 GPa) \cite{Liu2023}.  

Our calculations from the $r^2SCAN$ shows that the $Sm$-$type$ phase emerges as most stable phase in the pressure range of 9.2 and 18.6 GPa, followed by the transition into the dhcp phase after 18.6 GPa.  These enthalpy calculations using the $r^2SCAN$ functional show a close agreement with experimental measurements \cite{Samudrala2012,Grosshans1982}, highlighting the functional's efficacy in accurately predicting phase transitions. The enthalpy-pressure profile using $r^2SCAN$ is shown in Fig. \ref{fig:1}a, while same for other functionals are provided in the Supprting material (SM) as Fig. S1. To further validate the performance of the functionals, we analyzed the volume per yttrium (Y) atom. The Fig. \ref{fig:1}b indicates the volume calculated using the $r^2SCAN$ functional with compared to experimental data (indicated as star) \cite{Pace2020}. Similar results for other functionals are available in SM as Fig. S2. The $PBE$ functional exhibited a significant underestimation of the volume, indicating its limitations in accurately capturing the structural properties of yttrium under pressure. Conversely, $r^2SCAN$ and $SCAN$ reasonably provide better fit. The vdW correction fails to improve any results and drawing our conclusion that there is no major contribution of long-range forces in yttrium metal. There is 1.26 \% of volume drop at the first phase transition and the second phase transition is continuous.

Now turning our attention towards the understanding of phase transition mechanism for the yttrium based on static case. Earlier reports suggest that the reason behind the phase transition could be a charge transfer from the s to the d orbitals of the yttrium metal \cite{Duthie1977}. To confirm this behaviour, we performed a Mulliken charge population analysis using the LOBSTER package \cite{Nelson2020}. As shown in Table \ref{table2}, we observed that with increasing pressure, the 5s orbital is losing the charge while the $4d$ orbitals are gaining the charge. However, the $4s$ and $4p$ orbitals do not show significant changes over the pressure range in hcp phase. These results indicate an overall charge transfer from the s to d orbital and hence a modification in the interatomic interactions within the material which can further cause changes in the crystal lattice and leading to a phase transition. Interestingly, the high-pressure phases show nominal charge population changes in the $4s$ and $4p$ orbitals.  

\begin{table}[H] 
	\fontsize{10}{10}\selectfont
	\caption{\label{table2}%
		Analysis of Mulliken charge population for hcp, Sm-type, and dhcp phases under varying pressure conditions.
	}
	\centering
	\begin{ruledtabular}
		\begin{tabular}{lccccccc} 
			\multicolumn{7}{c}{\textbf{Values}} & 
			\textrm{\textbf{Trend}} \\
			\hline
			\multicolumn{8}{c}{\textbf{hcp}} \\ 
			\hline
			Pressure & \multirow{2}{*}{0.0001} & \multirow{2}{*}{2} & \multirow{2}{*}{4} & \multirow{2}{*}{6} & \multirow{2}{*}{8} & \multirow{2}{*}{10} & \\
			(GPa) &&&&&&& \\
			\textit{4s}  & 2.00 & 2.00 & 2.00 & 2.00 & 2.00 & 2.00 & =  \\
			\textit{5s}  & 1.31 & 1.29 & 1.26 & 1.24 & 1.22 & 1.20 & $\downarrow$  \\
			\textit{4p\textsubscript{y}} & 2.00 & 2.00 & 2.00 & 2.00 & 2.00 & 2.00 &  =  \\
			\textit{4p\textsubscript{z}} & 2.00 & 2.00 & 2.00 & 2.00 & 2.00 & 2.00 &  =  \\
			\textit{4p\textsubscript{x}} & 2.00 & 2.00 & 2.00 & 2.00 & 2.00 & 2.00 &  =  \\
			\textit{4d\textsubscript{xy}} & 0.34 & 0.34 & 0.34 & 0.35 & 0.35 & 0.36 & $\uparrow$  \\
			\textit{4d\textsubscript{yz}} & 0.33 & 0.33 & 0.33 & 0.34 & 0.34 & 0.35 & $\uparrow$  \\
			\textit{4d\textsubscript{xz}} & 0.33 & 0.33 & 0.33 & 0.34 & 0.34 & 0.35 & $\uparrow$  \\
			\textit{4d\textsubscript{z2}} & 0.38 & 0.39 & 0.40 & 0.40 & 0.40 & 0.41 & $\uparrow$  \\
			\textit{4d\textsubscript{x2-y2}} & 0.33 & 0.34 & 0.34 & 0.34 & 0.35 & 0.35 & $\uparrow$  \\
			\hline
			\multicolumn{8}{c}{\textbf{Sm-type}} \\ 
			\hline
			Pressure & \multirow{2}{*}{12} & \multirow{2}{*}{14} & \multirow{2}{*}{16} & \multirow{2}{*}{18} & \multirow{2}{*}{20} & \multirow{2}{*}{22} & \\
			(GPa) &&&&&&& \\
			\textit{4s}  & 2.00 & 2.00 & 2.00 & 2.01 & 2.01 & 2.01 & $\approx$ \\
			\textit{5s}  & 1.15 & 1.13 & 1.11 & 1.10 & 1.08 & 1.07 & $\downarrow$  \\
			\textit{4p\textsubscript{y}} & 2.00 & 2.00 & 2.00 & 1.99 & 1.99 & 1.99 &  $\approx$  \\
			\textit{4p\textsubscript{z}} & 2.00 & 1.99 & 1.99 & 1.99 & 1.99 & 1.99 &  $\approx$  \\
			\textit{4p\textsubscript{x}} & 2.00 & 2.00 & 2.00 & 1.99 & 1.99 & 1.99 &  $\approx$  \\
			\textit{4d\textsubscript{xy}} & 0.37 & 0.37 & 0.38 & 0.38 & 0.38 & 0.39 & $\uparrow$  \\
			\textit{4d\textsubscript{yz}} & 0.38 & 0.38 & 0.38 & 0.38 & 0.39 & 0.39 & $\uparrow$  \\
			\textit{4d\textsubscript{xz}} & 0.37 & 0.38 & 0.38 & 0.38 & 0.39 & 0.39 & $\uparrow$  \\
			\textit{4d\textsubscript{z2}} & 0.38 & 0.38 & 0.38 & 0.38 & 0.39 & 0.39 & $\uparrow$  \\
			\textit{4d\textsubscript{x2-y2}} & 0.37 & 0.37 & 0.38 & 0.38 & 0.38 & 0.39 & $\uparrow$  \\
			\hline
			\multicolumn{8}{c}{\textbf{dhcp}} \\ 
			\hline
			Pressure & \multirow{2}{*}{22} & \multirow{2}{*}{24} & \multirow{2}{*}{26} & \multirow{2}{*}{28} & \multirow{2}{*}{30} &  & \\
			(GPa) &&&&&&& \\
			\textit{4s}  & 2.01 & 2.01 & 2.01 & 2.02 & 2.02 & & $\approx$ \\
			\textit{5s}  & 1.08 & 1.07 & 1.06 & 1.04 & 1.03 & & $\downarrow$  \\
			\textit{4p\textsubscript{y}} & 1.99 & 1.99 & 1.99 & 1.99 & 1.99 & &  $\approx$  \\
			\textit{4p\textsubscript{z}} & 1.99 & 1.99 & 1.99 & 1.99 & 1.99 & &  $\approx$  \\
			\textit{4p\textsubscript{x}} & 1.99 & 1.99 & 1.99 & 1.99 & 1.99 & &  $\approx$  \\
			\textit{4d\textsubscript{xy}} & 0.38 & 0.39 & 0.39 & 0.40 & 0.40 & & $\uparrow$  \\
			\textit{4d\textsubscript{yz}} & 0.39 & 0.39 & 0.40 & 0.40 & 0.41 & & $\uparrow$  \\
			\textit{4d\textsubscript{xz}} & 0.37 & 0.37 & 0.37 & 0.37 & 0.38 & & $\uparrow$  \\
			\textit{4d\textsubscript{z2}} & 0.39 & 0.39 & 0.40 & 0.40 & 0.41 & & $\uparrow$  \\
			\textit{4d\textsubscript{x2-y2}} & 0.38 & 0.39 & 0.39 & 0.40 & 0.40 & & $\uparrow$  \\
		\end{tabular}
		\normalsize
	\end{ruledtabular}
\end{table}

\begin{figure}[h]
	\centering
	
	\begin{minipage}[b]{0.70\columnwidth}
		\raggedright
		{\fontsize{12}{14}\selectfont \textbf{(a)}} \\  
		\includegraphics[width=\linewidth]{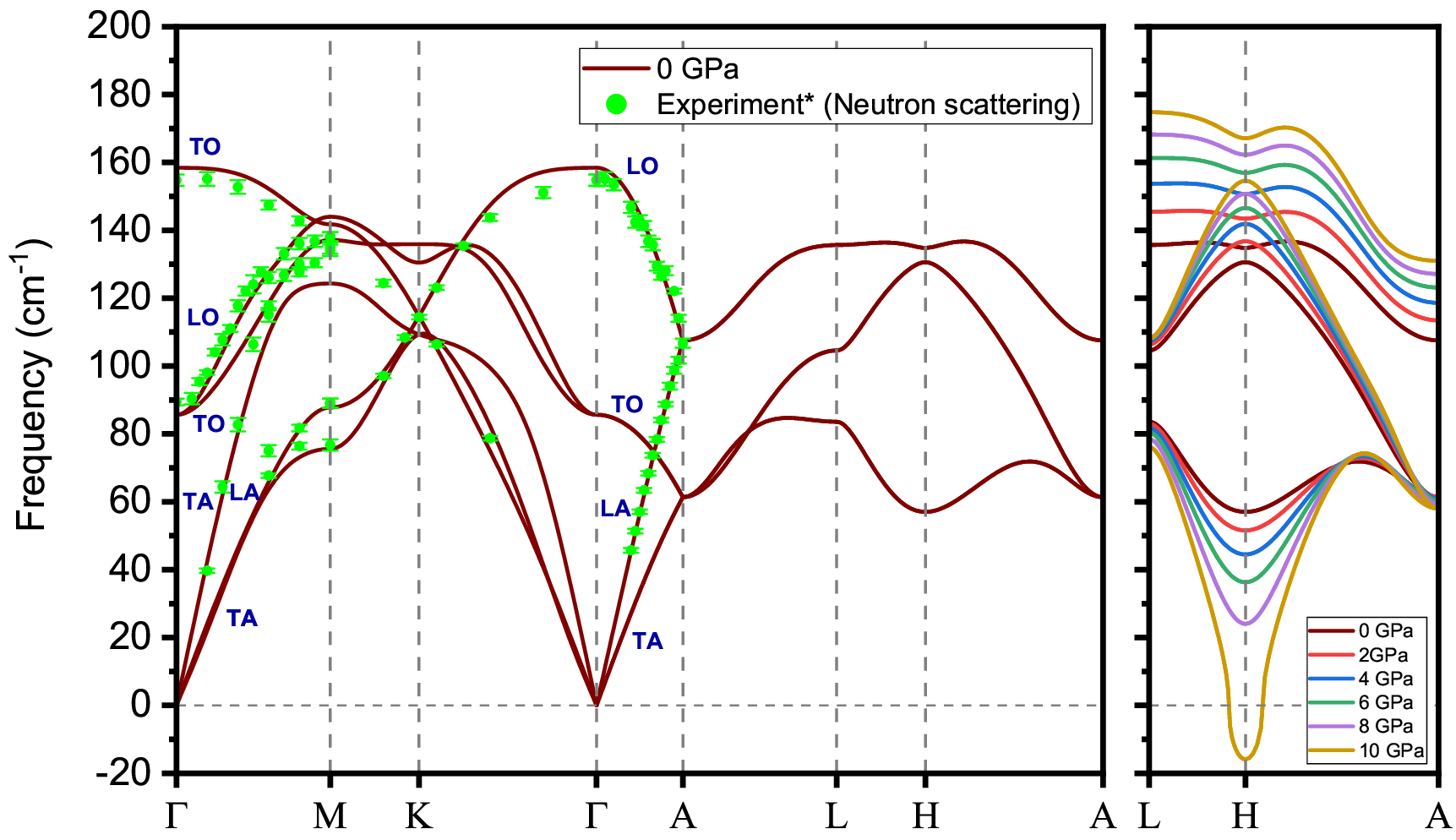}
	\end{minipage}
	\hspace{2pt} 
	\begin{minipage}[b]{0.17\columnwidth}
		\raggedright
		{\fontsize{12}{14}\selectfont \textbf{(b)}} \\  
		\includegraphics[width=\linewidth]{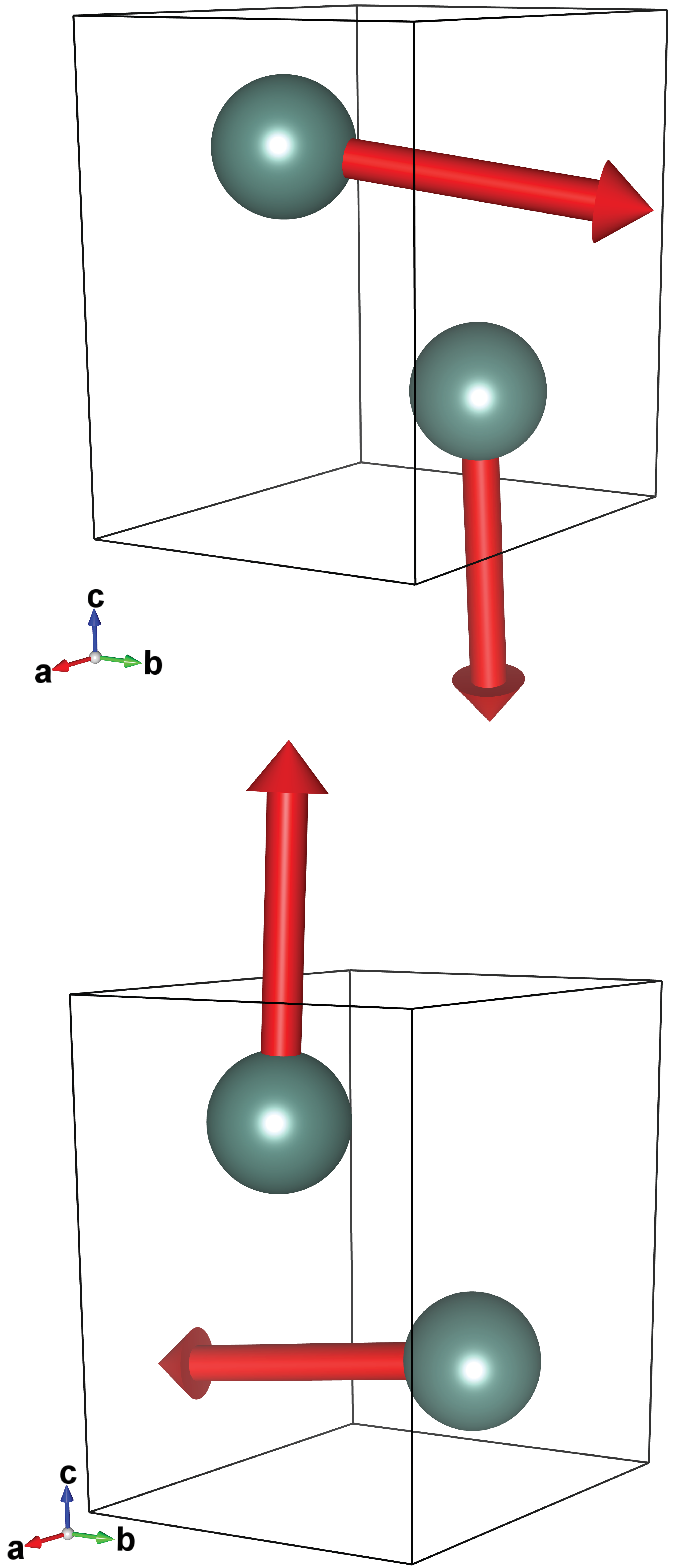}
	\end{minipage}
	
	\caption{Phonon dispersion curves of hcp phase at 1 atm pressure with comparison of *experimental data \cite{Sinha1970} indicated as green symbol with error bars along the L$\rightarrow$H$\rightarrow$A direction of BZ over the pressure range (a) and eigen vectors at H symmetry point at 10 GPa (b).}
	\label{fig:3}
\end{figure}

Apart from the static analysis, we also analysed dynamical and mechanical instabilities during $hcp$ to $Sm$-$type$ structural phase transition. The phonon dispersion curves calculated using the $r^2SCAN$ (Fig. \ref{fig:3}a) closely align with the available inelastic neutron scattering results \cite{Sinha1970}  for $hcp$ phase at 1 atm. The slight deviations observed are likely attributable to the temperature dependence of phonon spectra, a factor also noted in experimental studies. The transverse acoustic mode along the [001] ($\Gamma\rightarrow A$) direction is doubly degenerate. This degeneracy continues to persist along the high-symmetry path $A$$\rightarrow$L$\rightarrow$H$\rightarrow$$A$, where minor softening is observed at the edge of BZ (H-symmetry point). As the pressure increases, the amplitude of this softening increases and turns imaginary at the 10 GPa. The high amplitude of vibrations associated with this doubly degenerate mode leads to structural instability and phase transitions into $hcp$ to $Sm$-$type$. The direction of atomic vibrations associated with this imaginary phonon modes is shown in the Fig. \ref{fig:3}b. 

\begin{figure}[t]
	\centering
	
	\begin{minipage}{0.45\columnwidth} 
		\raggedright 
		{\fontsize{12}{14}\selectfont \textbf{(a)}} \\  
		\includegraphics[width=\linewidth]{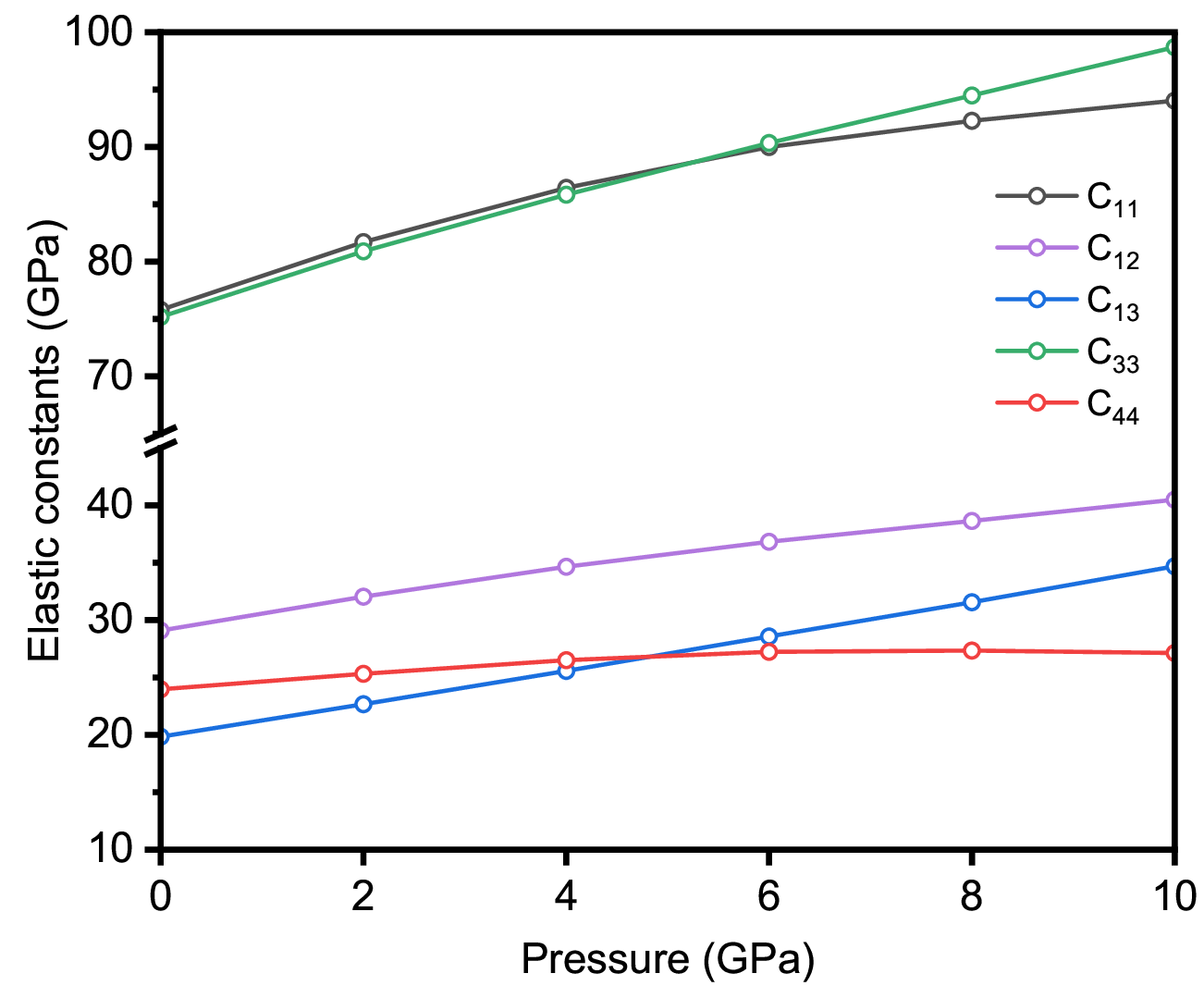} 
	\end{minipage}
	\hspace{2pt} 
	\begin{minipage}{0.45\columnwidth} 
		\raggedright 
		{\fontsize{12}{14}\selectfont \textbf{(b)}} \\  
		\includegraphics[width=\linewidth]{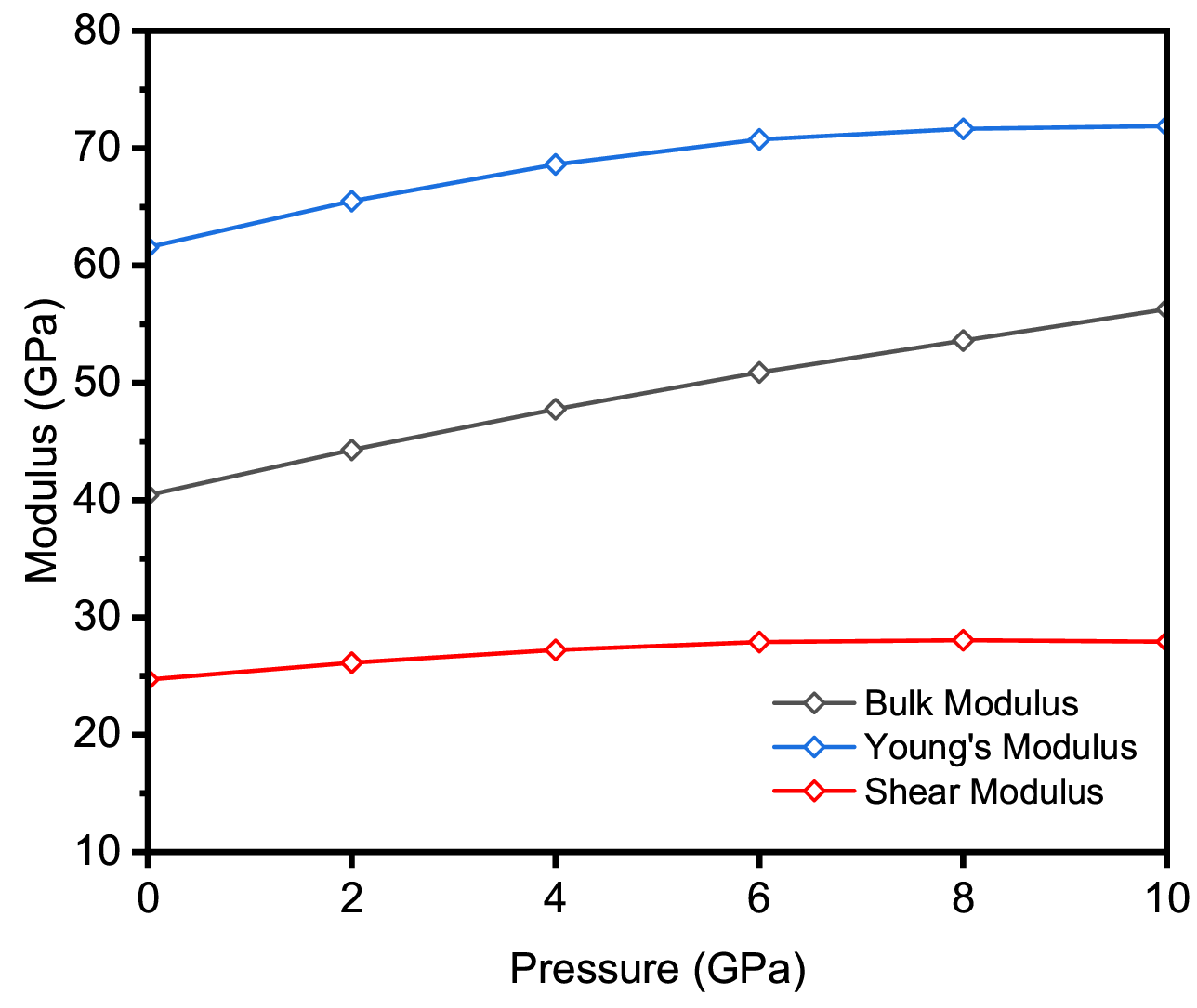} 
	\end{minipage}
	
	\caption{The elastic constants (a) and mechanical modulus (b) of hcp over the pressure range.}
	
	\label{fig:4}
\end{figure}

The softening in the acoustic mode often reflects in the elastic properties of material as the elastic constants are determined by analysing sound waves propagating in different directions, characterized by the slope ($\omega/q$) of the phonon dispersion curves. For the $hcp$ phase, there are five independent elastic constants: $C_{11}, C_{12}, C_{13}, C_{33},$ and $C_{44}$.Table \ref{table3} presents a comparative study of our $r^2SCAN$ calculations alongside previously reported theoretical results and experimental data on elastic properties at 1 atm pressure. The well agreement with the experimental data further demonstrates the reliability of our calculations. The pressure dependence of these elastic constants is shown in the Fig. \ref{fig:4}a. Here, the elastic constant $C_{44}$ softens after the 8 GPa correlating with the transverse acoustic modes and its softening in the phonon dispersions. The average mechanical properties, such as bulk modulus is related to $C_{11}, C_{12}, C_{13} $and $C_{33}$, while shear modulus is related to all of these constants. This relationship is also reflected in the values shown in Fig. \ref{fig:4}b. The longitudinal and transverse sound wave velocities (SM Fig. S3) further confirm that the softening of the transverse acoustic modes is responsible for the phase transition from $hcp$ to $Sm$-$type$.

\begin{table}[h] 
	\caption{\label{table3}%
		The elastic properties of hcp at 1 atm with comparison of experimental data.}
	\centering
	\begin{ruledtabular}
		\fontsize{10}{10}\selectfont
		\begin{tabular}{cccccc} 
			\multicolumn{1}{c}{\begin{tabular}[c]{@{}c@{}}Properties \\    (GPa)\end{tabular}} & \multicolumn{1}{c}{\begin{tabular}[c]{@{}c@{}}Present \\    (r2SCAN)\end{tabular}} & \multicolumn{1}{c}{\begin{tabular}[c]{@{}c@{}}Theory\footnotemark[1] \\ (PBE-\\ GGA)\end{tabular}} & \multicolumn{1}{c}{\begin{tabular}[c]{@{}c@{}}Theory\footnotemark[1]\\ (2NN\\ MEAM)\end{tabular}} & \multicolumn{1}{c}{\begin{tabular}[c]{@{}c@{}}Theory\footnotemark[2] \\ (Finnins-\\ Sinclair \\ interatomic \\ potential)\end{tabular}} & \multicolumn{1}{c}{\begin{tabular}[c]{@{}c@{}}Exp.\footnotemark[3]\\    (at 4.2 K)\end{tabular}} \\
			\colrule
			Bulk modulus & 40.42 & 43.34 & 44.88 & 44.70 & 41.17 \\
			$C_{11}$ & 75.80 & 80.40 & 77.74 & 70.86 & 83.40  \\
			$C_{12}$ & 29.09 & 17.06 & 30.00 & 34.55 & 29.10 \\
			$C_{13}$ & 19.83 & 15.85 & 28.21 & 29.40 & 19.00\\
			$C_{33}$ & 75.19 & 82.88 & 75.32 & 74.02 & 83.10 \\
			$C_{44}$ & 23.97 & 26.81 & 21.71 & 23.00 & 26.90 \\
		\end{tabular}
		\footnotetext[1]{Reference \cite{Ko2013}} 
		\footnotetext[2]{Reference \cite{Liu2023}}
		\footnotetext[3]{Reference \cite{Smith1960}}
		\normalsize
	\end{ruledtabular}
\end{table}

\begin{figure}[h]
	\centering
	
	\begin{minipage}[b]{0.80\columnwidth}
		\raggedright
		{\fontsize{12}{14}\selectfont \textbf{(a)}} \\  
		\includegraphics[width=\linewidth]{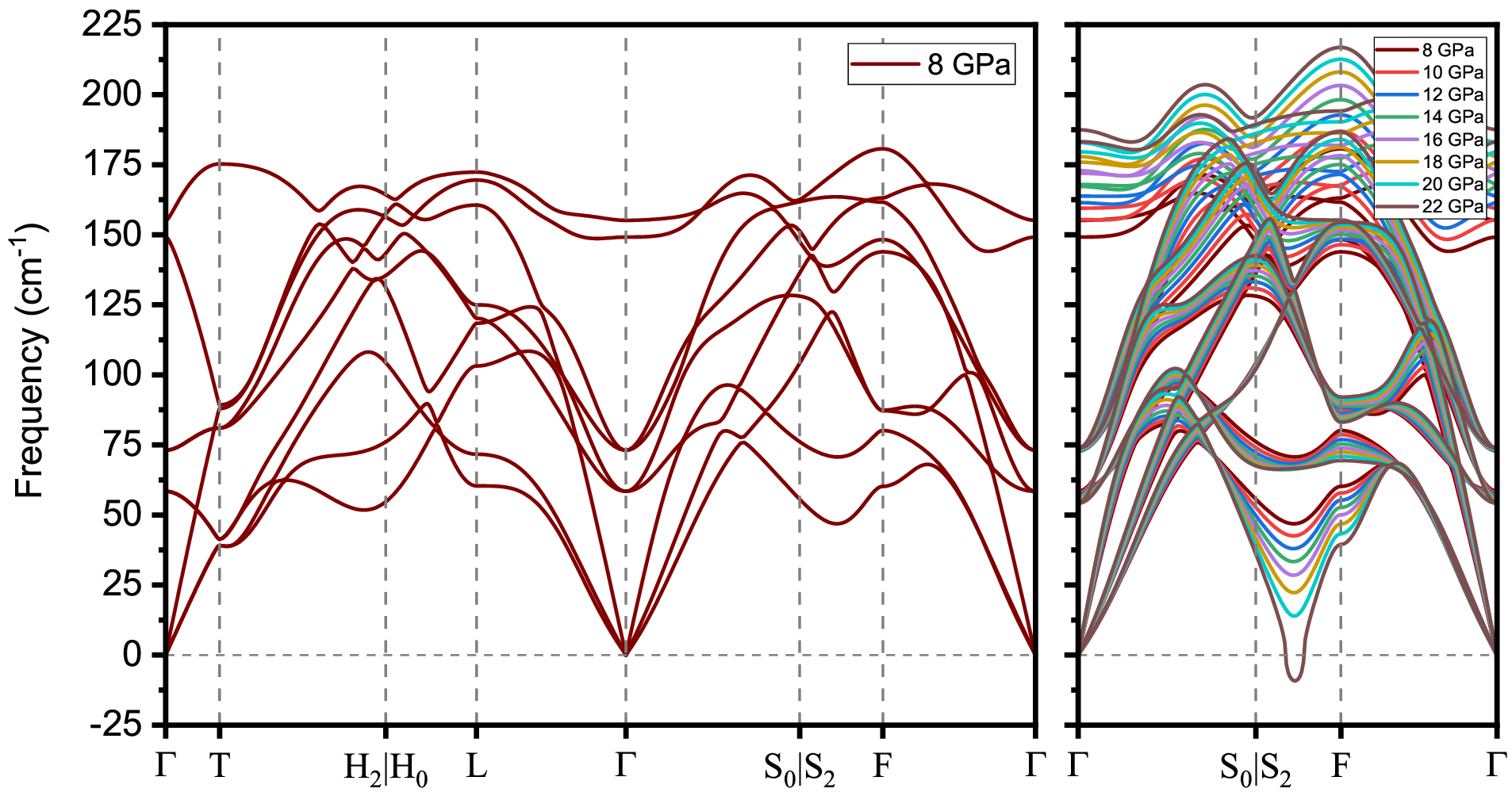}
	\end{minipage}
	\hspace{2pt} 
	\begin{minipage}[b]{0.086\columnwidth}
		\raggedright
		{\fontsize{12}{14}\selectfont \textbf{(b)}} \\  
		\includegraphics[width=\linewidth]{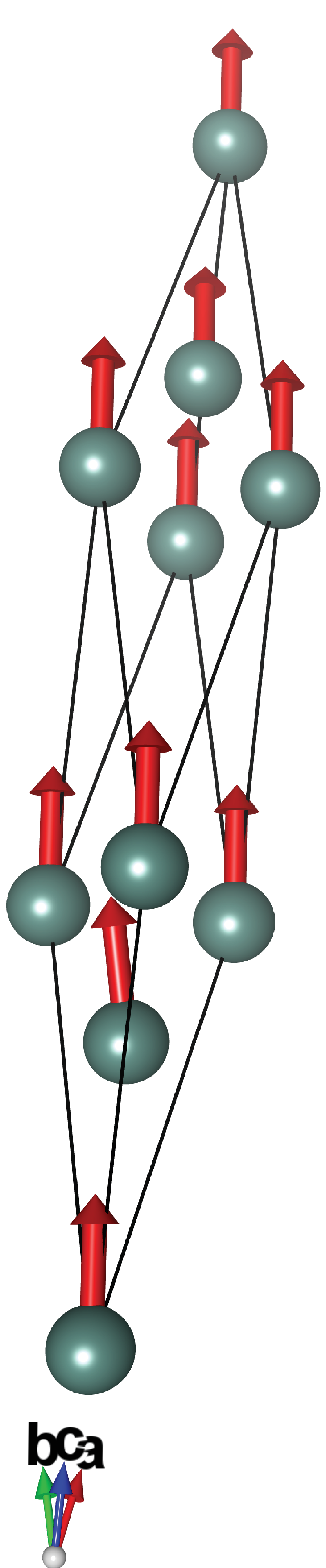}
	\end{minipage}
	
	\caption{Phonon dispersion curves of Sm-type phase at 8 GPa (a), along the $\Gamma \rightarrow S \rightarrow F \rightarrow \Gamma$ kpath over the pressure range and eigen vectors at mid $S_2-F$ point at 22 GPa (b).}
	\label{fig:5}
\end{figure}

The phonon dispersion curves have been calculated for the $Sm$-$type$ phase in the energetically stable pressure range to investigate whether a similar mechanism drives the subsequent phase transition. The phonon dispersion curves in the pressure range of 8 GPa to 22 GPa have been plotted in the Fig. \ref{fig:5}a which depict softness of the acoustic mode along the $S_2\rightarrow$$F$ path. This softening of the phonon mode leads to an instability and trigger the phase transition. The direction of vibrations leading to this imaginary mode is shown in the Fig. \ref{fig:5}b. Interestingly, the $Sm$-$type$ phase does not exhibit softening in its mechanical parameters, as shown in SI Fig. S4. This distinction emphasizes the complexity of phase transitions in yttrium and the interplay between vibrational and mechanical properties.

\FloatBarrier
\section{\label{IV}CONCLUSION}
In conclusion, the present comprehensive investigation focuses on yttrium metal upto 30 GPa pressure using first principles to predict correctly the phase transition pressure of the phases at lower pressure which has been of great concern due to its diverse values using different theoretical calculations and large deviations from experimental results. The present $r^2SCAN$ functional calculations produce much better than the earlier as well as present GGA calculations and in the good agreement with experimental values. The findings indicate that the phase transitions in yttrium are influenced by atomic vibrations together with the charge transfer. However, the primary reason for the phase transition could be vibrational instabilities as evidenced from the soft acoustic modes observed in the phonon dispersion curves of the $hcp$ and $Sm$-$type$ phases. Mechanical softening, particularly in the $hcp$ phase, is evident from elastic property calculations at the phase boundaries, indicating a close relationship between elastic instability and structural transitions. These findings point to the role of soft modes in phonon dispersion curves as a central factor in structural phase transitions of rare-earth materials.

\begin{acknowledgments}
Authors would like to acknowledge National Supercomputing Mission (NSM) for providing computing resource of ‘PARAM Porul’ at NIT Trichy, India. SP highly appreciates the Department of Science and Technology (DST), Government of India, for the INSPIRE fellowship (DST/INSPIRE Fellowship/2022/IF220593). This work is supported by the DST-FIST (SR/FST/PS/2022/230), Government of India. Part of the calculations were made at the Poznan Supercomputing and Networking Center (PSNC) as part of grant No. pl0331-01.
\end{acknowledgments}

\bibliographystyle{apsrev4-2} 
\bibliography{ref}

\begin{thebibliography}{37}%
\makeatletter
\providecommand \@ifxundefined [1]{%
 \@ifx{#1\undefined}
}%
\providecommand \@ifnum [1]{%
 \ifnum #1\expandafter \@firstoftwo
 \else \expandafter \@secondoftwo
 \fi
}%
\providecommand \@ifx [1]{%
 \ifx #1\expandafter \@firstoftwo
 \else \expandafter \@secondoftwo
 \fi
}%
\providecommand \natexlab [1]{#1}%
\providecommand \enquote  [1]{``#1''}%
\providecommand \bibnamefont  [1]{#1}%
\providecommand \bibfnamefont [1]{#1}%
\providecommand \citenamefont [1]{#1}%
\providecommand \href@noop [0]{\@secondoftwo}%
\providecommand \href [0]{\begingroup \@sanitize@url \@href}%
\providecommand \@href[1]{\@@startlink{#1}\@@href}%
\providecommand \@@href[1]{\endgroup#1\@@endlink}%
\providecommand \@sanitize@url [0]{\catcode `\\12\catcode `\$12\catcode
  `\&12\catcode `\#12\catcode `\^12\catcode `\_12\catcode `\%12\relax}%
\providecommand \@@startlink[1]{}%
\providecommand \@@endlink[0]{}%
\providecommand \url  [0]{\begingroup\@sanitize@url \@url }%
\providecommand \@url [1]{\endgroup\@href {#1}{\urlprefix }}%
\providecommand \urlprefix  [0]{URL }%
\providecommand \Eprint [0]{\href }%
\providecommand \doibase [0]{https://doi.org/}%
\providecommand \selectlanguage [0]{\@gobble}%
\providecommand \bibinfo  [0]{\@secondoftwo}%
\providecommand \bibfield  [0]{\@secondoftwo}%
\providecommand \translation [1]{[#1]}%
\providecommand \BibitemOpen [0]{}%
\providecommand \bibitemStop [0]{}%
\providecommand \bibitemNoStop [0]{.\EOS\space}%
\providecommand \EOS [0]{\spacefactor3000\relax}%
\providecommand \BibitemShut  [1]{\csname bibitem#1\endcsname}%
\let\auto@bib@innerbib\@empty
\bibitem [{\citenamefont {Benedict}\ \emph {et~al.}(1986)\citenamefont
  {Benedict}, \citenamefont {Grosshans},\ and\ \citenamefont
  {Holzapfel}}]{Benedict1986}%
  \BibitemOpen
  \bibfield  {author} {\bibinfo {author} {\bibfnamefont {U.}~\bibnamefont
  {Benedict}}, \bibinfo {author} {\bibfnamefont {W.~A.}\ \bibnamefont
  {Grosshans}},\ and\ \bibinfo {author} {\bibfnamefont {W.~B.}\ \bibnamefont
  {Holzapfel}},\ }\href {https://doi.org/10.1016/0378-4363(86)90283-4}
  {\bibfield  {journal} {\bibinfo  {journal} {Physica B+C}\ }\textbf {\bibinfo
  {volume} {144}},\ \bibinfo {pages} {14} (\bibinfo {year} {1986})}\BibitemShut
  {NoStop}%
\bibitem [{\citenamefont {Kr{\"{u}}ger}\ \emph {et~al.}(1990)\citenamefont
  {Kr{\"{u}}ger}, \citenamefont {Merkau}, \citenamefont {Grosshans},\ and\
  \citenamefont {Holzapfel}}]{Kruger1990}%
  \BibitemOpen
  \bibfield  {author} {\bibinfo {author} {\bibfnamefont {T.}~\bibnamefont
  {Kr{\"{u}}ger}}, \bibinfo {author} {\bibfnamefont {B.}~\bibnamefont
  {Merkau}}, \bibinfo {author} {\bibfnamefont {W.~A.}\ \bibnamefont
  {Grosshans}},\ and\ \bibinfo {author} {\bibfnamefont {W.~B.}\ \bibnamefont
  {Holzapfel}},\ }\href {https://doi.org/10.1080/08957959008203173} {\bibfield
  {journal} {\bibinfo  {journal} {High Pressure Research}\ }\textbf {\bibinfo
  {volume} {2}},\ \bibinfo {pages} {193} (\bibinfo {year} {1990})}\BibitemShut
  {NoStop}%
\bibitem [{\citenamefont {Grosshans}\ and\ \citenamefont
  {Holzapfel}(1992)}]{Grosshans1992}%
  \BibitemOpen
  \bibfield  {author} {\bibinfo {author} {\bibfnamefont {W.~A.}\ \bibnamefont
  {Grosshans}}\ and\ \bibinfo {author} {\bibfnamefont {W.~B.}\ \bibnamefont
  {Holzapfel}},\ }\href {https://doi.org/10.1103/PhysRevB.45.5171} {\bibfield
  {journal} {\bibinfo  {journal} {Physical Review B}\ }\textbf {\bibinfo
  {volume} {45}},\ \bibinfo {pages} {5171} (\bibinfo {year}
  {1992})}\BibitemShut {NoStop}%
\bibitem [{\citenamefont {Lei}\ \emph {et~al.}(2007)\citenamefont {Lei},
  \citenamefont {Papaconstantopoulos},\ and\ \citenamefont {Mehl}}]{Lei2007}%
  \BibitemOpen
  \bibfield  {author} {\bibinfo {author} {\bibfnamefont {S.}~\bibnamefont
  {Lei}}, \bibinfo {author} {\bibfnamefont {D.~A.}\ \bibnamefont
  {Papaconstantopoulos}},\ and\ \bibinfo {author} {\bibfnamefont {M.~J.}\
  \bibnamefont {Mehl}},\ }\href {https://doi.org/10.1103/PhysRevB.75.024512}
  {\bibfield  {journal} {\bibinfo  {journal} {Physical Review B}\ }\textbf
  {\bibinfo {volume} {75}},\ \bibinfo {pages} {024512} (\bibinfo {year}
  {2007})}\BibitemShut {NoStop}%
\bibitem [{\citenamefont {S{\"{o}}derlind}\ \emph {et~al.}(2014)\citenamefont
  {S{\"{o}}derlind}, \citenamefont {Turchi}, \citenamefont {Landa},\ and\
  \citenamefont {Lordi}}]{Soderlind2014}%
  \BibitemOpen
  \bibfield  {author} {\bibinfo {author} {\bibfnamefont {P.}~\bibnamefont
  {S{\"{o}}derlind}}, \bibinfo {author} {\bibfnamefont {P.~E.}\ \bibnamefont
  {Turchi}}, \bibinfo {author} {\bibfnamefont {A.}~\bibnamefont {Landa}},\ and\
  \bibinfo {author} {\bibfnamefont {V.}~\bibnamefont {Lordi}},\ }\href
  {https://doi.org/10.1088/0953-8984/26/41/416001} {\bibfield  {journal}
  {\bibinfo  {journal} {Journal of Physics Condensed Matter}\ }\textbf
  {\bibinfo {volume} {26}},\ \bibinfo {pages} {416001} (\bibinfo {year}
  {2014})}\BibitemShut {NoStop}%
\bibitem [{\citenamefont {Holzapfel}(1995)}]{Holzapfel1995}%
  \BibitemOpen
  \bibfield  {author} {\bibinfo {author} {\bibfnamefont {W.~B.}\ \bibnamefont
  {Holzapfel}},\ }\href {https://doi.org/10.1016/0925-8388(94)09001-7}
  {\bibfield  {journal} {\bibinfo  {journal} {Journal of Alloys and Compounds}\
  }\textbf {\bibinfo {volume} {223}},\ \bibinfo {pages} {170} (\bibinfo {year}
  {1995})}\BibitemShut {NoStop}%
\bibitem [{\citenamefont {McMahon}\ \emph {et~al.}(2019)\citenamefont
  {McMahon}, \citenamefont {Finnegan}, \citenamefont {Husband}, \citenamefont
  {Munro}, \citenamefont {Plekhanov}, \citenamefont {Bonini}, \citenamefont
  {Weber}, \citenamefont {Hanfland}, \citenamefont {Schwarz},\ and\
  \citenamefont {Macleod}}]{McMahon2019}%
  \BibitemOpen
  \bibfield  {author} {\bibinfo {author} {\bibfnamefont {M.~I.}\ \bibnamefont
  {McMahon}}, \bibinfo {author} {\bibfnamefont {S.}~\bibnamefont {Finnegan}},
  \bibinfo {author} {\bibfnamefont {R.~J.}\ \bibnamefont {Husband}}, \bibinfo
  {author} {\bibfnamefont {K.~A.}\ \bibnamefont {Munro}}, \bibinfo {author}
  {\bibfnamefont {E.}~\bibnamefont {Plekhanov}}, \bibinfo {author}
  {\bibfnamefont {N.}~\bibnamefont {Bonini}}, \bibinfo {author} {\bibfnamefont
  {C.}~\bibnamefont {Weber}}, \bibinfo {author} {\bibfnamefont
  {M.}~\bibnamefont {Hanfland}}, \bibinfo {author} {\bibfnamefont
  {U.}~\bibnamefont {Schwarz}},\ and\ \bibinfo {author} {\bibfnamefont {S.~G.}\
  \bibnamefont {Macleod}},\ }\href
  {https://doi.org/10.1103/PhysRevB.100.024107} {\bibfield  {journal} {\bibinfo
   {journal} {Physical Review B}\ }\textbf {\bibinfo {volume} {100}},\ \bibinfo
  {pages} {024107} (\bibinfo {year} {2019})}\BibitemShut {NoStop}%
\bibitem [{\citenamefont {Samudrala}\ \emph {et~al.}(2012)\citenamefont
  {Samudrala}, \citenamefont {Tsoi},\ and\ \citenamefont
  {Vohra}}]{Samudrala2012}%
  \BibitemOpen
  \bibfield  {author} {\bibinfo {author} {\bibfnamefont {G.~K.}\ \bibnamefont
  {Samudrala}}, \bibinfo {author} {\bibfnamefont {G.~M.}\ \bibnamefont
  {Tsoi}},\ and\ \bibinfo {author} {\bibfnamefont {Y.~K.}\ \bibnamefont
  {Vohra}},\ }\bibfield  {journal} {\bibinfo  {journal} {Journal of Physics
  Condensed Matter}\ }\textbf {\bibinfo {volume} {24}},\ \href
  {https://doi.org/10.1088/0953-8984/24/36/362201}
  {10.1088/0953-8984/24/36/362201} (\bibinfo {year} {2012})\BibitemShut
  {NoStop}%
\bibitem [{\citenamefont {Pace}\ \emph {et~al.}(2020)\citenamefont {Pace},
  \citenamefont {Finnegan}, \citenamefont {Storm}, \citenamefont {Stevenson},
  \citenamefont {McMahon}, \citenamefont {MacLeod}, \citenamefont {Plekhanov},
  \citenamefont {Bonini},\ and\ \citenamefont {Weber}}]{Pace2020}%
  \BibitemOpen
  \bibfield  {author} {\bibinfo {author} {\bibfnamefont {E.~J.}\ \bibnamefont
  {Pace}}, \bibinfo {author} {\bibfnamefont {S.~E.}\ \bibnamefont {Finnegan}},
  \bibinfo {author} {\bibfnamefont {C.~V.}\ \bibnamefont {Storm}}, \bibinfo
  {author} {\bibfnamefont {M.}~\bibnamefont {Stevenson}}, \bibinfo {author}
  {\bibfnamefont {M.~I.}\ \bibnamefont {McMahon}}, \bibinfo {author}
  {\bibfnamefont {S.~G.}\ \bibnamefont {MacLeod}}, \bibinfo {author}
  {\bibfnamefont {E.}~\bibnamefont {Plekhanov}}, \bibinfo {author}
  {\bibfnamefont {N.}~\bibnamefont {Bonini}},\ and\ \bibinfo {author}
  {\bibfnamefont {C.}~\bibnamefont {Weber}},\ }\href
  {https://doi.org/10.1103/PhysRevB.102.094104} {\bibfield  {journal} {\bibinfo
   {journal} {Physical Review B}\ }\textbf {\bibinfo {volume} {102}},\ \bibinfo
  {pages} {1} (\bibinfo {year} {2020})}\BibitemShut {NoStop}%
\bibitem [{\citenamefont {Johansson}\ and\ \citenamefont
  {Rosengren}(1975)}]{Johansson1975}%
  \BibitemOpen
  \bibfield  {author} {\bibinfo {author} {\bibfnamefont {B.}~\bibnamefont
  {Johansson}}\ and\ \bibinfo {author} {\bibfnamefont {A.}~\bibnamefont
  {Rosengren}},\ }\href {https://doi.org/10.1103/PhysRevB.11.2836} {\bibfield
  {journal} {\bibinfo  {journal} {Physical Review B}\ }\textbf {\bibinfo
  {volume} {11}},\ \bibinfo {pages} {2836} (\bibinfo {year}
  {1975})}\BibitemShut {NoStop}%
\bibitem [{\citenamefont {Chesnut}\ and\ \citenamefont
  {Vohra}(1998)}]{Chesnut1998}%
  \BibitemOpen
  \bibfield  {author} {\bibinfo {author} {\bibfnamefont {G.~N.}\ \bibnamefont
  {Chesnut}}\ and\ \bibinfo {author} {\bibfnamefont {Y.~K.}\ \bibnamefont
  {Vohra}},\ }\href {https://doi.org/10.1103/PhysRevB.57.10221} {\bibfield
  {journal} {\bibinfo  {journal} {Physical Review B - Condensed Matter and
  Materials Physics}\ }\textbf {\bibinfo {volume} {57}},\ \bibinfo {pages}
  {10221} (\bibinfo {year} {1998})}\BibitemShut {NoStop}%
\bibitem [{\citenamefont {Patterson}\ \emph {et~al.}(2004)\citenamefont
  {Patterson}, \citenamefont {Saw},\ and\ \citenamefont
  {Akella}}]{Patterson2004}%
  \BibitemOpen
  \bibfield  {author} {\bibinfo {author} {\bibfnamefont {R.}~\bibnamefont
  {Patterson}}, \bibinfo {author} {\bibfnamefont {C.~K.}\ \bibnamefont {Saw}},\
  and\ \bibinfo {author} {\bibfnamefont {J.}~\bibnamefont {Akella}},\ }\href
  {https://doi.org/10.1063/1.1699489} {\bibfield  {journal} {\bibinfo
  {journal} {Journal of Applied Physics}\ }\textbf {\bibinfo {volume} {95}},\
  \bibinfo {pages} {5443} (\bibinfo {year} {2004})}\BibitemShut {NoStop}%
\bibitem [{\citenamefont {Samudrala}\ \emph {et~al.}(2011)\citenamefont
  {Samudrala}, \citenamefont {Thomas}, \citenamefont {Montgomery},\ and\
  \citenamefont {Vohra}}]{Samudrala2011}%
  \BibitemOpen
  \bibfield  {author} {\bibinfo {author} {\bibfnamefont {G.~K.}\ \bibnamefont
  {Samudrala}}, \bibinfo {author} {\bibfnamefont {S.~A.}\ \bibnamefont
  {Thomas}}, \bibinfo {author} {\bibfnamefont {J.~M.}\ \bibnamefont
  {Montgomery}},\ and\ \bibinfo {author} {\bibfnamefont {Y.~K.}\ \bibnamefont
  {Vohra}},\ }\bibfield  {journal} {\bibinfo  {journal} {Journal of Physics
  Condensed Matter}\ }\textbf {\bibinfo {volume} {23}},\ \href
  {https://doi.org/10.1088/0953-8984/23/31/315701}
  {10.1088/0953-8984/23/31/315701} (\bibinfo {year} {2011})\BibitemShut
  {NoStop}%
\bibitem [{\citenamefont {Duthie}\ and\ \citenamefont
  {Pettifor}(1977)}]{Duthie1977}%
  \BibitemOpen
  \bibfield  {author} {\bibinfo {author} {\bibfnamefont {J.~C.}\ \bibnamefont
  {Duthie}}\ and\ \bibinfo {author} {\bibfnamefont {D.~G.}\ \bibnamefont
  {Pettifor}},\ }\href {https://doi.org/10.1103/PhysRevLett.38.564} {\bibfield
  {journal} {\bibinfo  {journal} {Physical Review Letters}\ }\textbf {\bibinfo
  {volume} {38}},\ \bibinfo {pages} {564} (\bibinfo {year} {1977})}\BibitemShut
  {NoStop}%
\bibitem [{\citenamefont {Vohra}\ \emph {et~al.}(1981)\citenamefont {Vohra},
  \citenamefont {Olijnik}, \citenamefont {Grosshans},\ and\ \citenamefont
  {Holzapfel}}]{Vohra1981}%
  \BibitemOpen
  \bibfield  {author} {\bibinfo {author} {\bibfnamefont {Y.~K.}\ \bibnamefont
  {Vohra}}, \bibinfo {author} {\bibfnamefont {H.}~\bibnamefont {Olijnik}},
  \bibinfo {author} {\bibfnamefont {W.}~\bibnamefont {Grosshans}},\ and\
  \bibinfo {author} {\bibfnamefont {W.~B.}\ \bibnamefont {Holzapfel}},\ }\href
  {https://doi.org/10.1103/PhysRevLett.47.1065} {\bibfield  {journal} {\bibinfo
   {journal} {Physical Review Letters}\ }\textbf {\bibinfo {volume} {47}},\
  \bibinfo {pages} {1065} (\bibinfo {year} {1981})}\BibitemShut {NoStop}%
\bibitem [{\citenamefont {Grosshans}\ \emph {et~al.}(1982)\citenamefont
  {Grosshans}, \citenamefont {Vohra},\ and\ \citenamefont
  {Holzapfel}}]{Grosshans1982}%
  \BibitemOpen
  \bibfield  {author} {\bibinfo {author} {\bibfnamefont {W.~A.}\ \bibnamefont
  {Grosshans}}, \bibinfo {author} {\bibfnamefont {Y.~K.}\ \bibnamefont
  {Vohra}},\ and\ \bibinfo {author} {\bibfnamefont {W.~B.}\ \bibnamefont
  {Holzapfel}},\ }\href {https://doi.org/10.1016/0304-8853(82)90251-7}
  {\bibfield  {journal} {\bibinfo  {journal} {Journal of Magnetism and Magnetic
  Materials}\ }\textbf {\bibinfo {volume} {29}},\ \bibinfo {pages} {282}
  (\bibinfo {year} {1982})}\BibitemShut {NoStop}%
\bibitem [{\citenamefont {Chen}\ \emph {et~al.}(2011)\citenamefont {Chen},
  \citenamefont {Hu},\ and\ \citenamefont {Yang}}]{Chen2011}%
  \BibitemOpen
  \bibfield  {author} {\bibinfo {author} {\bibfnamefont {Y.}~\bibnamefont
  {Chen}}, \bibinfo {author} {\bibfnamefont {Q.-M.}\ \bibnamefont {Hu}},\ and\
  \bibinfo {author} {\bibfnamefont {R.}~\bibnamefont {Yang}},\ }\href
  {https://doi.org/10.1103/PhysRevB.84.132101} {\bibfield  {journal} {\bibinfo
  {journal} {Physical Review B}\ }\textbf {\bibinfo {volume} {84}},\ \bibinfo
  {pages} {132101} (\bibinfo {year} {2011})}\BibitemShut {NoStop}%
\bibitem [{\citenamefont {Buhot}\ \emph {et~al.}(2020)\citenamefont {Buhot},
  \citenamefont {Moulding}, \citenamefont {Muramatsu}, \citenamefont {Osmond},\
  and\ \citenamefont {Friedemann}}]{Buhot2020}%
  \BibitemOpen
  \bibfield  {author} {\bibinfo {author} {\bibfnamefont {J.}~\bibnamefont
  {Buhot}}, \bibinfo {author} {\bibfnamefont {O.}~\bibnamefont {Moulding}},
  \bibinfo {author} {\bibfnamefont {T.}~\bibnamefont {Muramatsu}}, \bibinfo
  {author} {\bibfnamefont {I.}~\bibnamefont {Osmond}},\ and\ \bibinfo {author}
  {\bibfnamefont {S.}~\bibnamefont {Friedemann}},\ }\href
  {https://doi.org/10.1103/PhysRevB.102.104508} {\bibfield  {journal} {\bibinfo
   {journal} {Physical Review B}\ }\textbf {\bibinfo {volume} {102}},\ \bibinfo
  {pages} {1} (\bibinfo {year} {2020})},\ \Eprint
  {https://arxiv.org/abs/2007.03969} {arXiv:2007.03969} \BibitemShut {NoStop}%
\bibitem [{\citenamefont {Chen}\ \emph {et~al.}(2012)\citenamefont {Chen},
  \citenamefont {Hu},\ and\ \citenamefont {Yang}}]{Chen2012}%
  \BibitemOpen
  \bibfield  {author} {\bibinfo {author} {\bibfnamefont {Y.}~\bibnamefont
  {Chen}}, \bibinfo {author} {\bibfnamefont {Q.~M.}\ \bibnamefont {Hu}},\ and\
  \bibinfo {author} {\bibfnamefont {R.}~\bibnamefont {Yang}},\ }\href
  {https://doi.org/10.1103/PhysRevLett.109.157004} {\bibfield  {journal}
  {\bibinfo  {journal} {Physical Review Letters}\ }\textbf {\bibinfo {volume}
  {109}},\ \bibinfo {pages} {157004} (\bibinfo {year} {2012})}\BibitemShut
  {NoStop}%
\bibitem [{\citenamefont {Hamlin}\ \emph {et~al.}(2007)\citenamefont {Hamlin},
  \citenamefont {Tissen},\ and\ \citenamefont {Schilling}}]{Hamlin2007}%
  \BibitemOpen
  \bibfield  {author} {\bibinfo {author} {\bibfnamefont {J.~J.}\ \bibnamefont
  {Hamlin}}, \bibinfo {author} {\bibfnamefont {V.~G.}\ \bibnamefont {Tissen}},\
  and\ \bibinfo {author} {\bibfnamefont {J.~S.}\ \bibnamefont {Schilling}},\
  }\href {https://doi.org/10.1016/j.physc.2006.10.012} {\bibfield  {journal}
  {\bibinfo  {journal} {Physica C: Superconductivity and its Applications}\
  }\textbf {\bibinfo {volume} {451}},\ \bibinfo {pages} {82} (\bibinfo {year}
  {2007})}\BibitemShut {NoStop}%
\bibitem [{\citenamefont {Li}\ \emph {et~al.}(2019)\citenamefont {Li},
  \citenamefont {Mei}, \citenamefont {Lu}, \citenamefont {Xiang}, \citenamefont
  {Zhang}, \citenamefont {Du}, \citenamefont {Wang},\ and\ \citenamefont
  {Chen}}]{Li2019}%
  \BibitemOpen
  \bibfield  {author} {\bibinfo {author} {\bibfnamefont {P.}~\bibnamefont
  {Li}}, \bibinfo {author} {\bibfnamefont {T.}~\bibnamefont {Mei}}, \bibinfo
  {author} {\bibfnamefont {Z.}~\bibnamefont {Lu}}, \bibinfo {author}
  {\bibfnamefont {L.}~\bibnamefont {Xiang}}, \bibinfo {author} {\bibfnamefont
  {X.}~\bibnamefont {Zhang}}, \bibinfo {author} {\bibfnamefont
  {X.}~\bibnamefont {Du}}, \bibinfo {author} {\bibfnamefont {J.}~\bibnamefont
  {Wang}},\ and\ \bibinfo {author} {\bibfnamefont {H.}~\bibnamefont {Chen}},\
  }\href {https://doi.org/10.1016/j.commatsci.2018.12.022} {\bibfield
  {journal} {\bibinfo  {journal} {Computational Materials Science}\ }\textbf
  {\bibinfo {volume} {159}},\ \bibinfo {pages} {428} (\bibinfo {year}
  {2019})}\BibitemShut {NoStop}%
\bibitem [{\citenamefont {Giannessi}\ \emph {et~al.}(2024)\citenamefont
  {Giannessi}, \citenamefont {{Di Cataldo}}, \citenamefont {Saha},\ and\
  \citenamefont {Boeri}}]{Giannessi2024}%
  \BibitemOpen
  \bibfield  {author} {\bibinfo {author} {\bibfnamefont {F.}~\bibnamefont
  {Giannessi}}, \bibinfo {author} {\bibfnamefont {S.}~\bibnamefont {{Di
  Cataldo}}}, \bibinfo {author} {\bibfnamefont {S.}~\bibnamefont {Saha}},\ and\
  \bibinfo {author} {\bibfnamefont {L.}~\bibnamefont {Boeri}},\ }\href
  {https://doi.org/10.1038/s41597-024-03447-1} {\bibfield  {journal} {\bibinfo
  {journal} {Scientific Data}\ }\textbf {\bibinfo {volume} {11}},\ \bibinfo
  {pages} {1} (\bibinfo {year} {2024})}\BibitemShut {NoStop}%
\bibitem [{\citenamefont {Liu}\ \emph {et~al.}(2023)\citenamefont {Liu},
  \citenamefont {Chen}, \citenamefont {Guo}, \citenamefont {Li}, \citenamefont
  {Wang}, \citenamefont {Deng}, \citenamefont {Tian}, \citenamefont {Hu},
  \citenamefont {Xiao},\ and\ \citenamefont {Yuan}}]{Liu2023}%
  \BibitemOpen
  \bibfield  {author} {\bibinfo {author} {\bibfnamefont {B.}~\bibnamefont
  {Liu}}, \bibinfo {author} {\bibfnamefont {Y.}~\bibnamefont {Chen}}, \bibinfo
  {author} {\bibfnamefont {L.}~\bibnamefont {Guo}}, \bibinfo {author}
  {\bibfnamefont {X.}~\bibnamefont {Li}}, \bibinfo {author} {\bibfnamefont
  {K.}~\bibnamefont {Wang}}, \bibinfo {author} {\bibfnamefont {H.}~\bibnamefont
  {Deng}}, \bibinfo {author} {\bibfnamefont {Z.}~\bibnamefont {Tian}}, \bibinfo
  {author} {\bibfnamefont {W.}~\bibnamefont {Hu}}, \bibinfo {author}
  {\bibfnamefont {S.}~\bibnamefont {Xiao}},\ and\ \bibinfo {author}
  {\bibfnamefont {D.}~\bibnamefont {Yuan}},\ }\bibfield  {journal} {\bibinfo
  {journal} {International Journal of Mechanical Sciences}\ }\textbf {\bibinfo
  {volume} {250}},\ \href {https://doi.org/10.1016/j.ijmecsci.2023.108330}
  {10.1016/j.ijmecsci.2023.108330} (\bibinfo {year} {2023})\BibitemShut
  {NoStop}%
\bibitem [{\citenamefont {Li}\ \emph {et~al.}(2022)\citenamefont {Li},
  \citenamefont {Zhuang}, \citenamefont {Li}, \citenamefont {Ou}, \citenamefont
  {Feng}, \citenamefont {Zhang}, \citenamefont {Bai}, \citenamefont {Ma},
  \citenamefont {Jin},\ and\ \citenamefont {Liu}}]{Li2022}%
  \BibitemOpen
  \bibfield  {author} {\bibinfo {author} {\bibfnamefont {X.}~\bibnamefont
  {Li}}, \bibinfo {author} {\bibfnamefont {Q.}~\bibnamefont {Zhuang}}, \bibinfo
  {author} {\bibfnamefont {D.}~\bibnamefont {Li}}, \bibinfo {author}
  {\bibfnamefont {T.}~\bibnamefont {Ou}}, \bibinfo {author} {\bibfnamefont
  {S.}~\bibnamefont {Feng}}, \bibinfo {author} {\bibfnamefont {J.}~\bibnamefont
  {Zhang}}, \bibinfo {author} {\bibfnamefont {W.}~\bibnamefont {Bai}}, \bibinfo
  {author} {\bibfnamefont {X.}~\bibnamefont {Ma}}, \bibinfo {author}
  {\bibfnamefont {X.}~\bibnamefont {Jin}},\ and\ \bibinfo {author}
  {\bibfnamefont {J.}~\bibnamefont {Liu}},\ }\href
  {https://doi.org/10.1016/j.ssc.2022.115004} {\bibfield  {journal} {\bibinfo
  {journal} {Solid State Communications}\ }\textbf {\bibinfo {volume} {358}},\
  \bibinfo {pages} {115004} (\bibinfo {year} {2022})}\BibitemShut {NoStop}%
\bibitem [{\citenamefont {Dalsaniya}\ \emph {et~al.}(2022)\citenamefont
  {Dalsaniya}, \citenamefont {Kurzyd{\l}owski},\ and\ \citenamefont
  {Kurzyd{\l}owski}}]{Dalsaniya2022}%
  \BibitemOpen
  \bibfield  {author} {\bibinfo {author} {\bibfnamefont {M.~H.}\ \bibnamefont
  {Dalsaniya}}, \bibinfo {author} {\bibfnamefont {K.~J.}\ \bibnamefont
  {Kurzyd{\l}owski}},\ and\ \bibinfo {author} {\bibfnamefont {D.}~\bibnamefont
  {Kurzyd{\l}owski}},\ }\href {https://doi.org/10.1103/PhysRevB.106.115128}
  {\bibfield  {journal} {\bibinfo  {journal} {Physical Review B}\ }\textbf
  {\bibinfo {volume} {106}},\ \bibinfo {pages} {115128} (\bibinfo {year}
  {2022})}\BibitemShut {NoStop}%
\bibitem [{\citenamefont {Yang}\ \emph {et~al.}(2020)\citenamefont {Yang},
  \citenamefont {Liu}, \citenamefont {Lu},\ and\ \citenamefont
  {Lin}}]{Yang2020}%
  \BibitemOpen
  \bibfield  {author} {\bibinfo {author} {\bibfnamefont {H.~C.}\ \bibnamefont
  {Yang}}, \bibinfo {author} {\bibfnamefont {K.}~\bibnamefont {Liu}}, \bibinfo
  {author} {\bibfnamefont {Z.~Y.}\ \bibnamefont {Lu}},\ and\ \bibinfo {author}
  {\bibfnamefont {H.~Q.}\ \bibnamefont {Lin}},\ }\href
  {https://doi.org/10.1103/PhysRevB.102.174109} {\bibfield  {journal} {\bibinfo
   {journal} {Physical Review B}\ }\textbf {\bibinfo {volume} {102}},\ \bibinfo
  {pages} {174109} (\bibinfo {year} {2020})}\BibitemShut {NoStop}%
\bibitem [{\citenamefont {Perdew}\ \emph {et~al.}(1996)\citenamefont {Perdew},
  \citenamefont {Burke},\ and\ \citenamefont {Ernzerhof}}]{Perdew1996}%
  \BibitemOpen
  \bibfield  {author} {\bibinfo {author} {\bibfnamefont {J.~P.}\ \bibnamefont
  {Perdew}}, \bibinfo {author} {\bibfnamefont {K.}~\bibnamefont {Burke}},\ and\
  \bibinfo {author} {\bibfnamefont {M.}~\bibnamefont {Ernzerhof}},\ }\href
  {https://doi.org/10.1103/PhysRevLett.77.3865} {\bibfield  {journal} {\bibinfo
   {journal} {Physical Review Letters}\ }\textbf {\bibinfo {volume} {77}},\
  \bibinfo {pages} {3865} (\bibinfo {year} {1996})}\BibitemShut {NoStop}%
\bibitem [{\citenamefont {Sun}\ \emph {et~al.}(2015)\citenamefont {Sun},
  \citenamefont {Ruzsinszky},\ and\ \citenamefont {Perdew}}]{Sun2015}%
  \BibitemOpen
  \bibfield  {author} {\bibinfo {author} {\bibfnamefont {J.}~\bibnamefont
  {Sun}}, \bibinfo {author} {\bibfnamefont {A.}~\bibnamefont {Ruzsinszky}},\
  and\ \bibinfo {author} {\bibfnamefont {J.}~\bibnamefont {Perdew}},\ }\href
  {https://doi.org/10.1103/PhysRevLett.115.036402} {\bibfield  {journal}
  {\bibinfo  {journal} {Physical Review Letters}\ }\textbf {\bibinfo {volume}
  {115}},\ \bibinfo {pages} {1} (\bibinfo {year} {2015})},\ \Eprint
  {https://arxiv.org/abs/1504.03028} {arXiv:1504.03028} \BibitemShut {NoStop}%
\bibitem [{\citenamefont {Furness}\ \emph {et~al.}(2020)\citenamefont
  {Furness}, \citenamefont {Kaplan}, \citenamefont {Ning}, \citenamefont
  {Perdew},\ and\ \citenamefont {Sun}}]{Furness2020}%
  \BibitemOpen
  \bibfield  {author} {\bibinfo {author} {\bibfnamefont {J.~W.}\ \bibnamefont
  {Furness}}, \bibinfo {author} {\bibfnamefont {A.~D.}\ \bibnamefont {Kaplan}},
  \bibinfo {author} {\bibfnamefont {J.}~\bibnamefont {Ning}}, \bibinfo {author}
  {\bibfnamefont {J.~P.}\ \bibnamefont {Perdew}},\ and\ \bibinfo {author}
  {\bibfnamefont {J.}~\bibnamefont {Sun}},\ }\href
  {https://doi.org/10.1021/acs.jpclett.0c02405} {\bibfield  {journal} {\bibinfo
   {journal} {Journal of Physical Chemistry Letters}\ }\textbf {\bibinfo
  {volume} {11}},\ \bibinfo {pages} {8208} (\bibinfo {year} {2020})},\ \Eprint
  {https://arxiv.org/abs/2008.03374} {arXiv:2008.03374} \BibitemShut {NoStop}%
\bibitem [{\citenamefont {Kresse}\ and\ \citenamefont
  {Furthm{\"{u}}ller}(1996)}]{Kresse1996}%
  \BibitemOpen
  \bibfield  {author} {\bibinfo {author} {\bibfnamefont {G.}~\bibnamefont
  {Kresse}}\ and\ \bibinfo {author} {\bibfnamefont {J.}~\bibnamefont
  {Furthm{\"{u}}ller}},\ }\href {https://doi.org/10.1016/0927-0256(96)00008-0}
  {\bibfield  {journal} {\bibinfo  {journal} {Computational Materials Science}\
  }\textbf {\bibinfo {volume} {6}},\ \bibinfo {pages} {15} (\bibinfo {year}
  {1996})}\BibitemShut {NoStop}%
\bibitem [{\citenamefont {Bl{\"{o}}chl}(1994)}]{Blochl1994}%
  \BibitemOpen
  \bibfield  {author} {\bibinfo {author} {\bibfnamefont {P.~E.}\ \bibnamefont
  {Bl{\"{o}}chl}},\ }\href {https://doi.org/10.1103/PhysRevB.50.17953}
  {\bibfield  {journal} {\bibinfo  {journal} {Physical Review B}\ }\textbf
  {\bibinfo {volume} {50}},\ \bibinfo {pages} {17953} (\bibinfo {year}
  {1994})}\BibitemShut {NoStop}%
\bibitem [{\citenamefont {Grimme}\ \emph {et~al.}(2011)\citenamefont {Grimme},
  \citenamefont {Ehrlich},\ and\ \citenamefont {Goerigk}}]{Grimme2011}%
  \BibitemOpen
  \bibfield  {author} {\bibinfo {author} {\bibfnamefont {S.}~\bibnamefont
  {Grimme}}, \bibinfo {author} {\bibfnamefont {S.}~\bibnamefont {Ehrlich}},\
  and\ \bibinfo {author} {\bibfnamefont {L.}~\bibnamefont {Goerigk}},\ }\href
  {https://doi.org/10.1002/jcc.21759} {\bibfield  {journal} {\bibinfo
  {journal} {Journal of Computational Chemistry}\ }\textbf {\bibinfo {volume}
  {32}},\ \bibinfo {pages} {1456} (\bibinfo {year} {2011})}\BibitemShut
  {NoStop}%
\bibitem [{\citenamefont {Nelson}\ \emph {et~al.}(2020)\citenamefont {Nelson},
  \citenamefont {Ertural}, \citenamefont {George}, \citenamefont {Deringer},
  \citenamefont {Hautier},\ and\ \citenamefont {Dronskowski}}]{Nelson2020}%
  \BibitemOpen
  \bibfield  {author} {\bibinfo {author} {\bibfnamefont {R.}~\bibnamefont
  {Nelson}}, \bibinfo {author} {\bibfnamefont {C.}~\bibnamefont {Ertural}},
  \bibinfo {author} {\bibfnamefont {J.}~\bibnamefont {George}}, \bibinfo
  {author} {\bibfnamefont {V.~L.}\ \bibnamefont {Deringer}}, \bibinfo {author}
  {\bibfnamefont {G.}~\bibnamefont {Hautier}},\ and\ \bibinfo {author}
  {\bibfnamefont {R.}~\bibnamefont {Dronskowski}},\ }\href
  {https://doi.org/10.1002/jcc.26353} {\bibfield  {journal} {\bibinfo
  {journal} {Journal of Computational Chemistry}\ }\textbf {\bibinfo {volume}
  {41}},\ \bibinfo {pages} {1931} (\bibinfo {year} {2020})}\BibitemShut
  {NoStop}%
\bibitem [{\citenamefont {Momma}\ and\ \citenamefont
  {Izumi}(2011)}]{Momma2011}%
  \BibitemOpen
  \bibfield  {author} {\bibinfo {author} {\bibfnamefont {K.}~\bibnamefont
  {Momma}}\ and\ \bibinfo {author} {\bibfnamefont {F.}~\bibnamefont {Izumi}},\
  }\href {https://doi.org/10.1107/S0021889811038970} {\bibfield  {journal}
  {\bibinfo  {journal} {Journal of Applied Crystallography}\ }\textbf {\bibinfo
  {volume} {44}},\ \bibinfo {pages} {1272} (\bibinfo {year}
  {2011})}\BibitemShut {NoStop}%
\bibitem [{\citenamefont {Sinha}\ \emph {et~al.}(1970)\citenamefont {Sinha},
  \citenamefont {Brun}, \citenamefont {Muhlestein},\ and\ \citenamefont
  {Sakurai}}]{Sinha1970}%
  \BibitemOpen
  \bibfield  {author} {\bibinfo {author} {\bibfnamefont {S.~K.}\ \bibnamefont
  {Sinha}}, \bibinfo {author} {\bibfnamefont {T.~O.}\ \bibnamefont {Brun}},
  \bibinfo {author} {\bibfnamefont {L.~D.}\ \bibnamefont {Muhlestein}},\ and\
  \bibinfo {author} {\bibfnamefont {J.}~\bibnamefont {Sakurai}},\ }\href
  {https://doi.org/10.1103/PhysRevB.1.2430} {\bibfield  {journal} {\bibinfo
  {journal} {Physical Review B}\ }\textbf {\bibinfo {volume} {1}},\ \bibinfo
  {pages} {2430} (\bibinfo {year} {1970})}\BibitemShut {NoStop}%
\bibitem [{\citenamefont {Ko}\ and\ \citenamefont {Lee}(2013)}]{Ko2013}%
  \BibitemOpen
  \bibfield  {author} {\bibinfo {author} {\bibfnamefont {W.~S.}\ \bibnamefont
  {Ko}}\ and\ \bibinfo {author} {\bibfnamefont {B.~J.}\ \bibnamefont {Lee}},\
  }\bibfield  {journal} {\bibinfo  {journal} {Modelling and Simulation in
  Materials Science and Engineering}\ }\textbf {\bibinfo {volume} {21}},\ \href
  {https://doi.org/10.1088/0965-0393/21/8/085008}
  {10.1088/0965-0393/21/8/085008} (\bibinfo {year} {2013})\BibitemShut
  {NoStop}%
\bibitem [{\citenamefont {Smith}\ and\ \citenamefont
  {Gjevre}(1960)}]{Smith1960}%
  \BibitemOpen
  \bibfield  {author} {\bibinfo {author} {\bibfnamefont {J.~F.}\ \bibnamefont
  {Smith}}\ and\ \bibinfo {author} {\bibfnamefont {J.~A.}\ \bibnamefont
  {Gjevre}},\ }\href {https://doi.org/10.1063/1.1735657} {\bibfield  {journal}
  {\bibinfo  {journal} {Journal of Applied Physics}\ }\textbf {\bibinfo
  {volume} {31}},\ \bibinfo {pages} {645} (\bibinfo {year} {1960})}\BibitemShut
  {NoStop}%
\end{thebibliography}%

\end{document}


\title{\textbf{Supplemental Material \\ for \\ $``$Revisiting Phase Transition of Yttrium: \\ Insights from Density Functional Theory$"$}}

\author{Paras Patel$^{1}$, Madhavi H. Dalsaniya$^{2,3}$, Saurav Patel$^{1}$, Dominik Kurzydłowski$^{3}$, Krzysztof J. Kurzydłowski$^{2,4}$, Prafulla K. Jha$^{1,*}$}

\affiliation{$^1$Department of Physics, Faculty of Science, The Maharaja Sayajirao University of Baroda, Vadodara-390002, Gujarat, India.}
\affiliation{$^2$Faculty of Materials Science and Engineering, Warsaw University of Technology, Wołoska 141, 02-507, Warsaw, Poland.}
\affiliation{$^3$Faculty of Mathematics and Natural Sciences, Cardinal Stefan Wyszyński University in Warsaw, 01-038 Warsaw, Poland.}
\affiliation{$^4$Faculty of Mechanical Engineering, Bialystok University of Technology, Wiejska 45C, 15-351, Bialystok, Poland.}

\affiliation{$^*$Corresponding author: \textit{prafullaj@yahoo.com}}

\maketitle
\section{\label{I}Supporting Figures}

\begin{figure}[h]
	\centering
	
	\begin{minipage}[t]{0.32\textwidth}
		\raggedright
		{\fontsize{8}{12}\selectfont \textbf{(a)}} \\  
		\includegraphics[width=\linewidth]{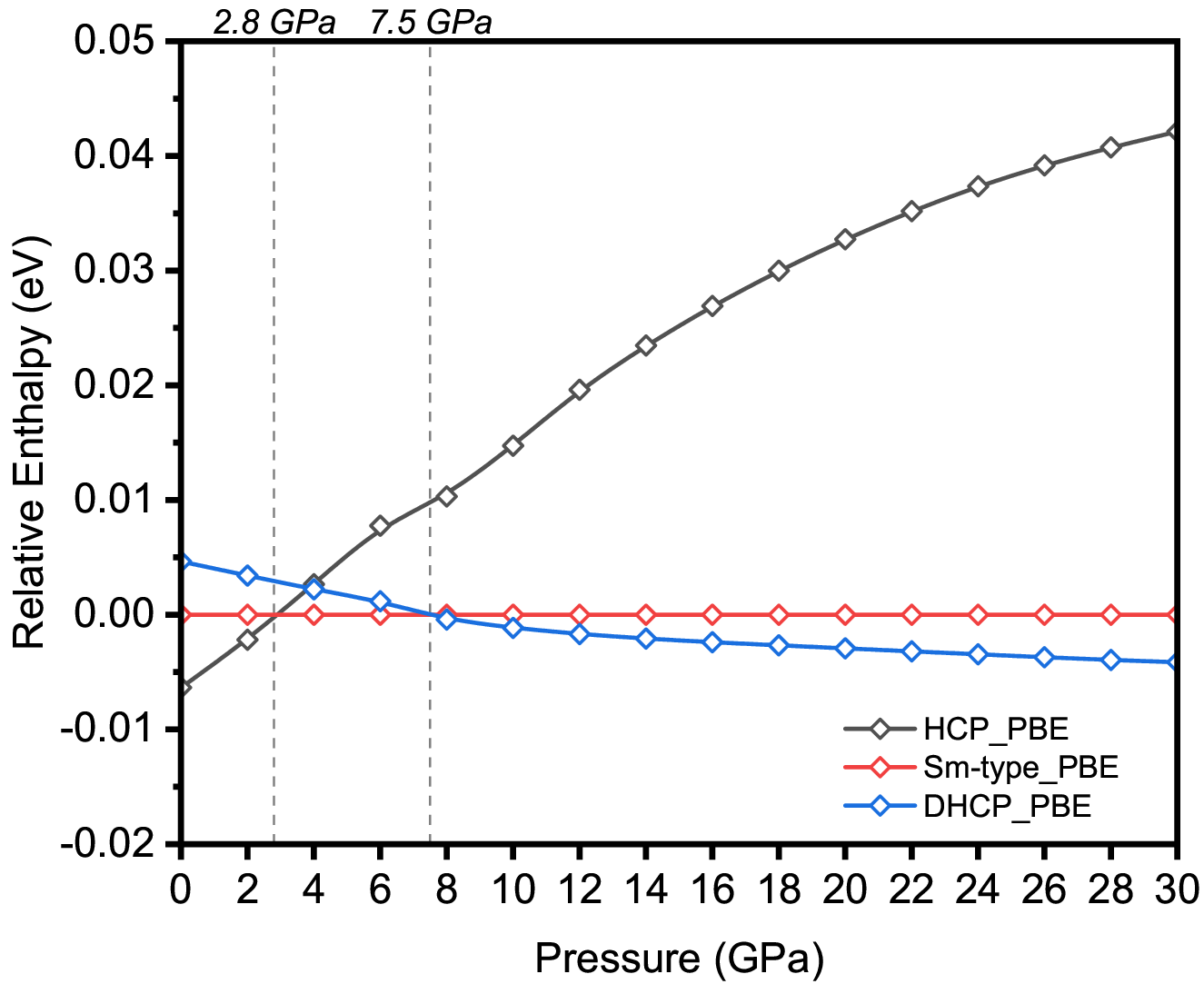}
	\end{minipage}
	\hfill
	\begin{minipage}[t]{0.32\textwidth}
		\raggedright
		{\fontsize{8}{12}\selectfont \textbf{(b)}} \\  
		\includegraphics[width=\linewidth]{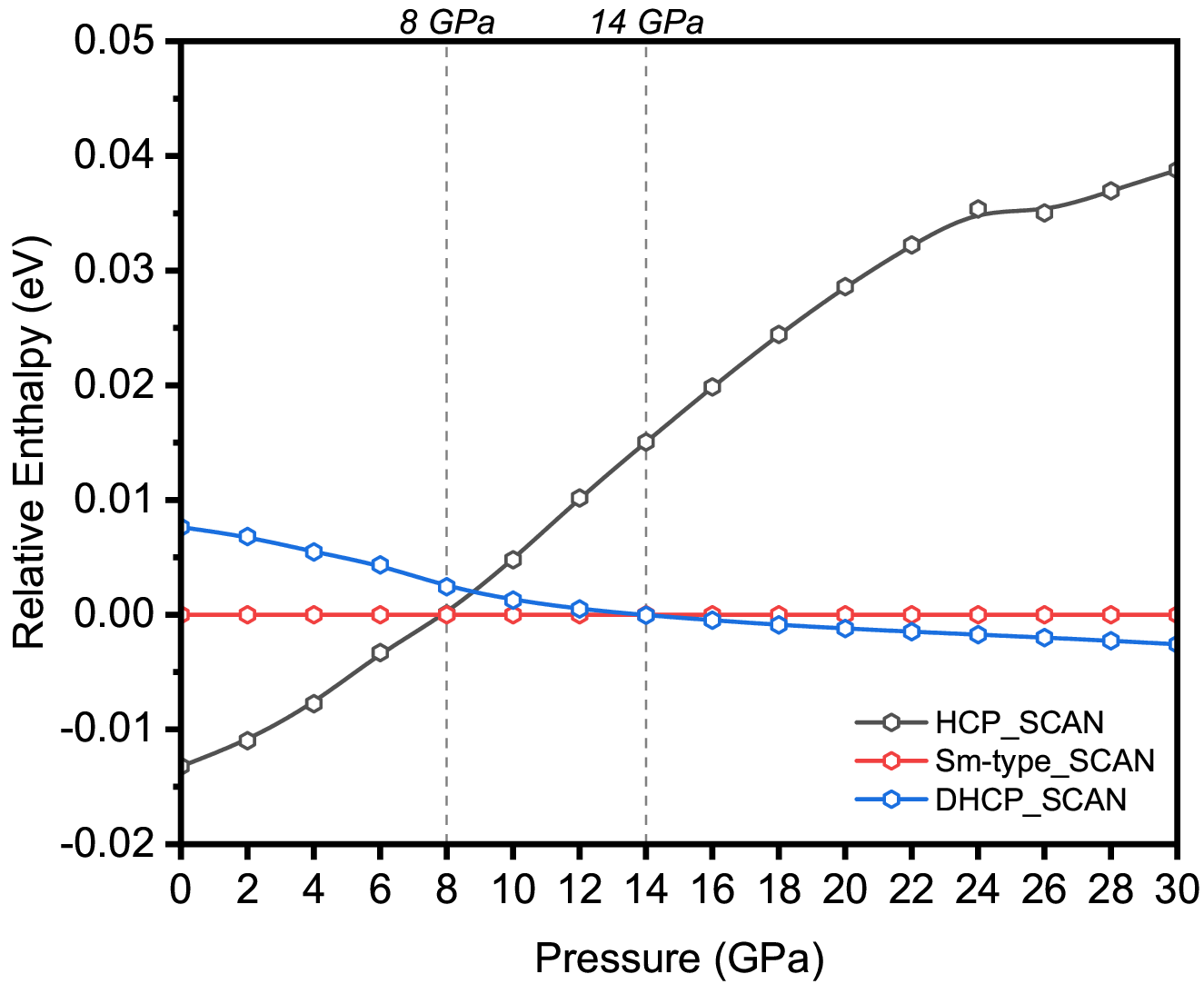}
	\end{minipage}
	\hfill
	\begin{minipage}[t]{0.32\textwidth}
		\raggedright
		{\fontsize{8}{12}\selectfont \textbf{(c)}} \\  
		\includegraphics[width=\linewidth]{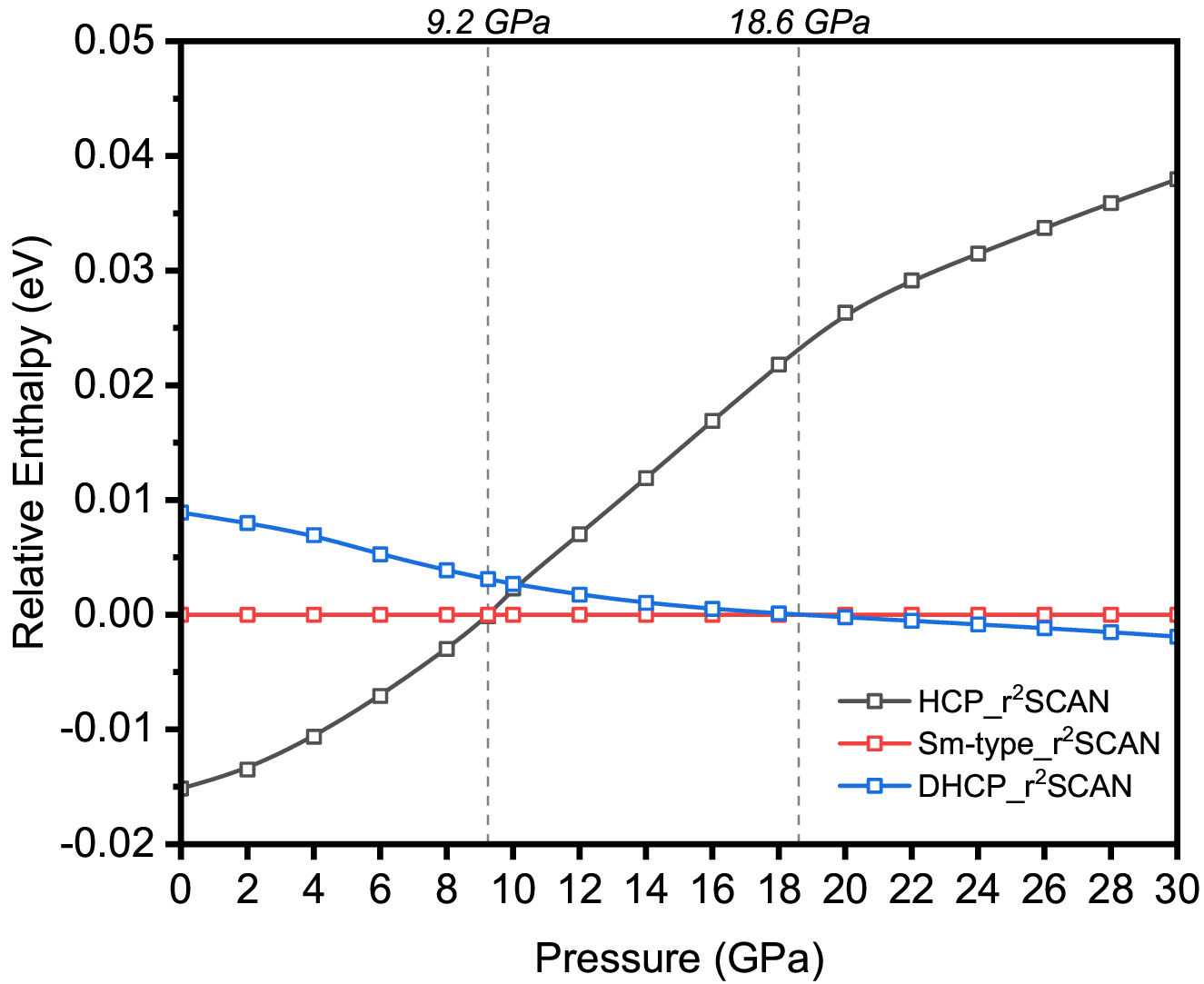}
	\end{minipage}
	
	\vspace{10pt}
	
	\begin{minipage}[t]{0.32\textwidth}
		\raggedright
		{\fontsize{8}{12}\selectfont \textbf{(d)}} \\  
		\includegraphics[width=\linewidth]{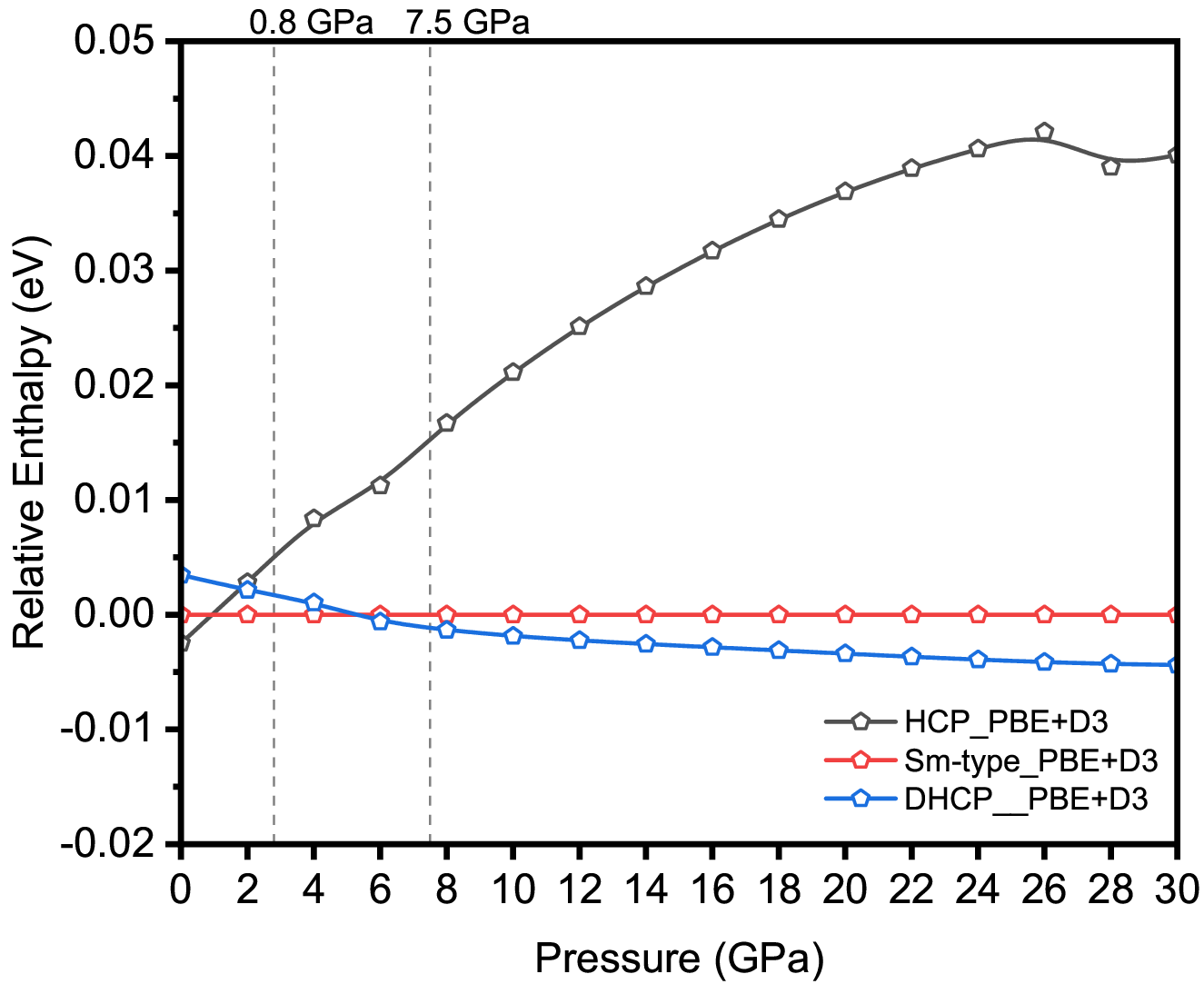}
	\end{minipage}
	\hspace{10pt} 
	\begin{minipage}[t]{0.32\textwidth}
		\raggedright
		{\fontsize{8}{12}\selectfont \textbf{(e)}} \\  
		\includegraphics[width=\linewidth]{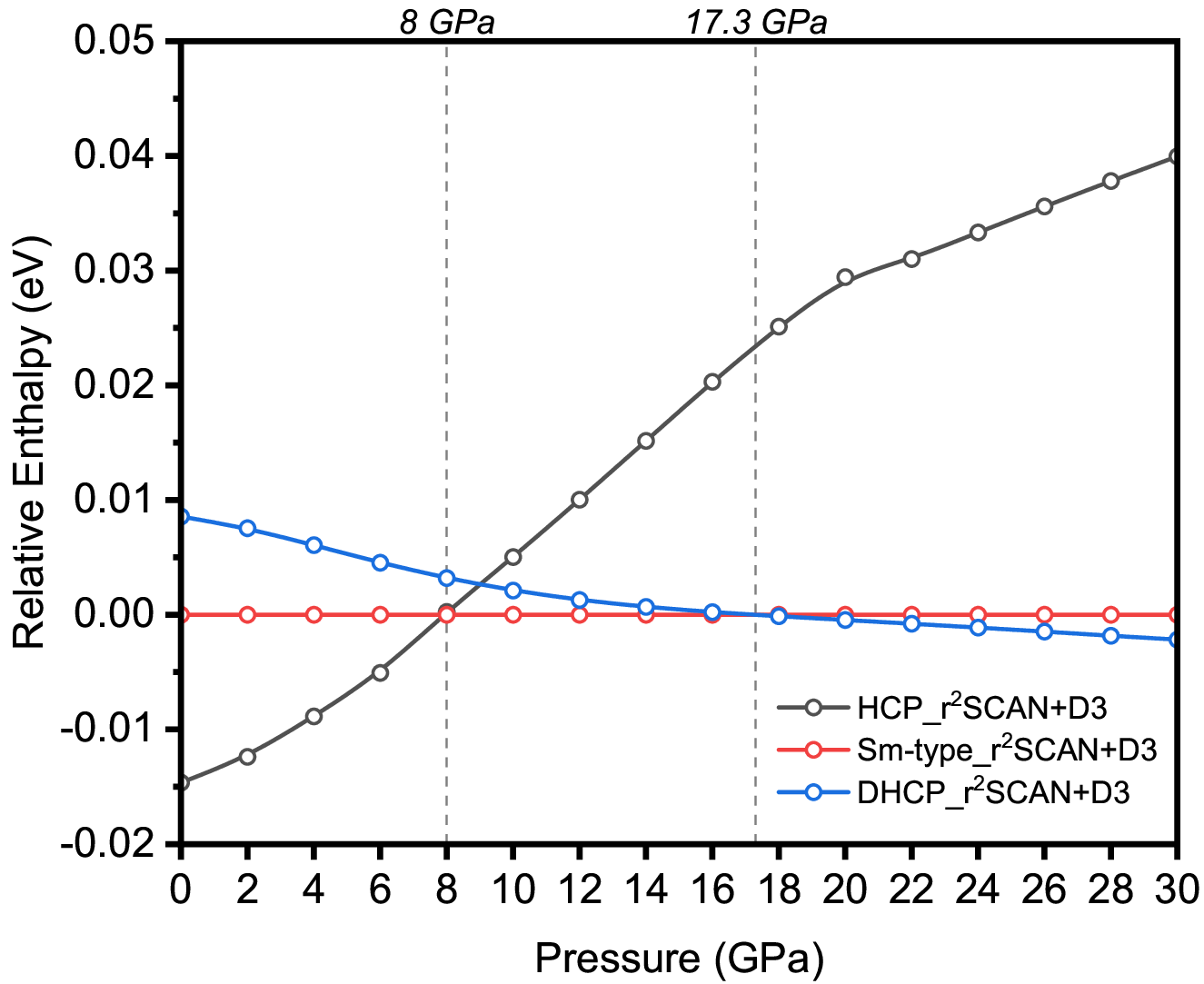}
	\end{minipage}
	
	\caption{The enthalpy-pressure profile using PBE-GGA (a), SCAN-metaGGA (b), r$^2$SCAN-metaGGA (c), and with dispersion-corrected D3 level of functionals (d) and (e).}
	\label{fig:S1}
\end{figure}

\begin{figure}[h]
	\centering
	
	\begin{minipage}[t]{0.32\textwidth}
		\raggedright
		{\fontsize{8}{12}\selectfont \textbf{(a)}} \\  
		\includegraphics[width=\linewidth]{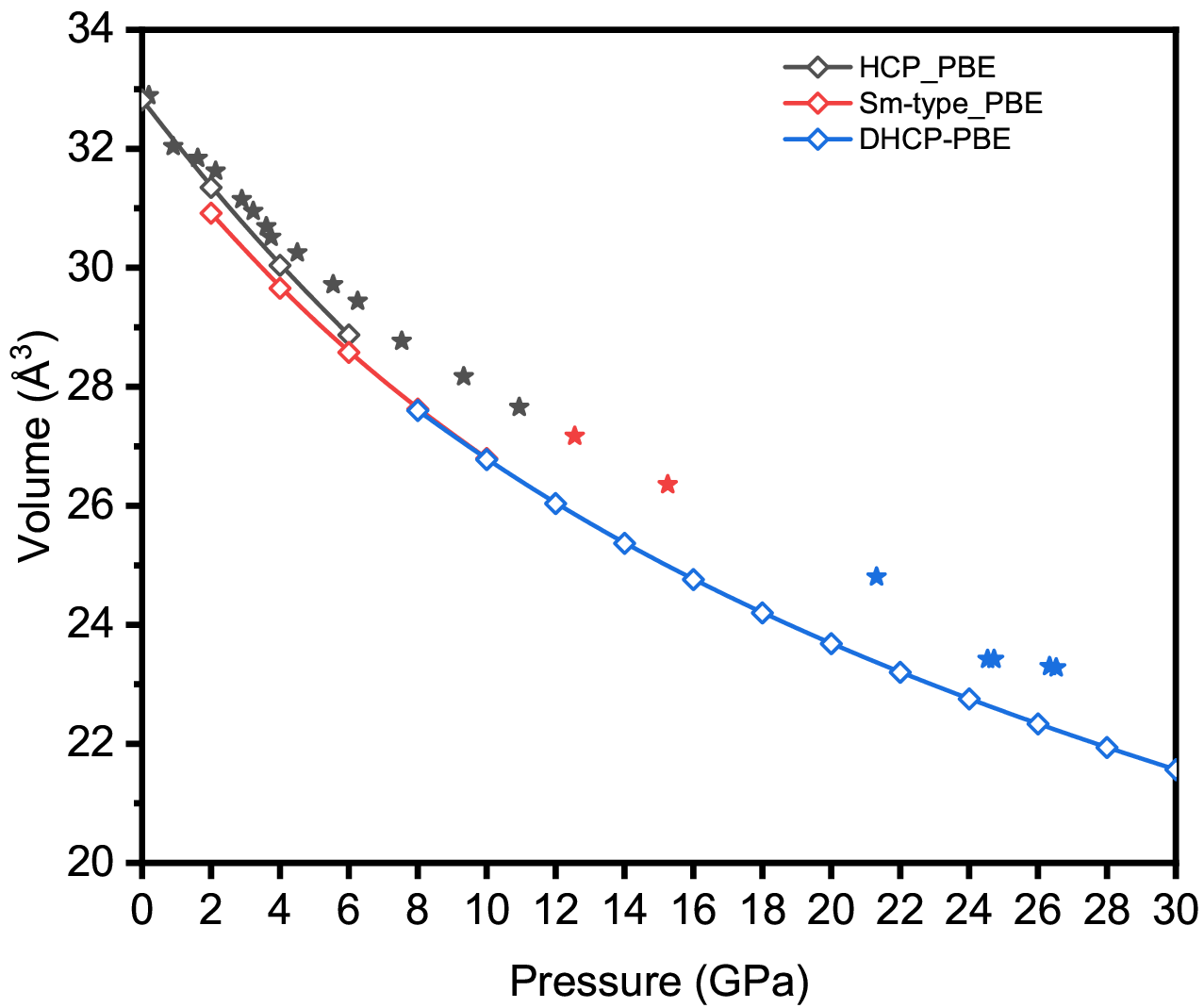}
	\end{minipage}
	\hfill
	\begin{minipage}[t]{0.32\textwidth}
		\raggedright
		{\fontsize{8}{12}\selectfont \textbf{(b)}} \\  
		\includegraphics[width=\linewidth]{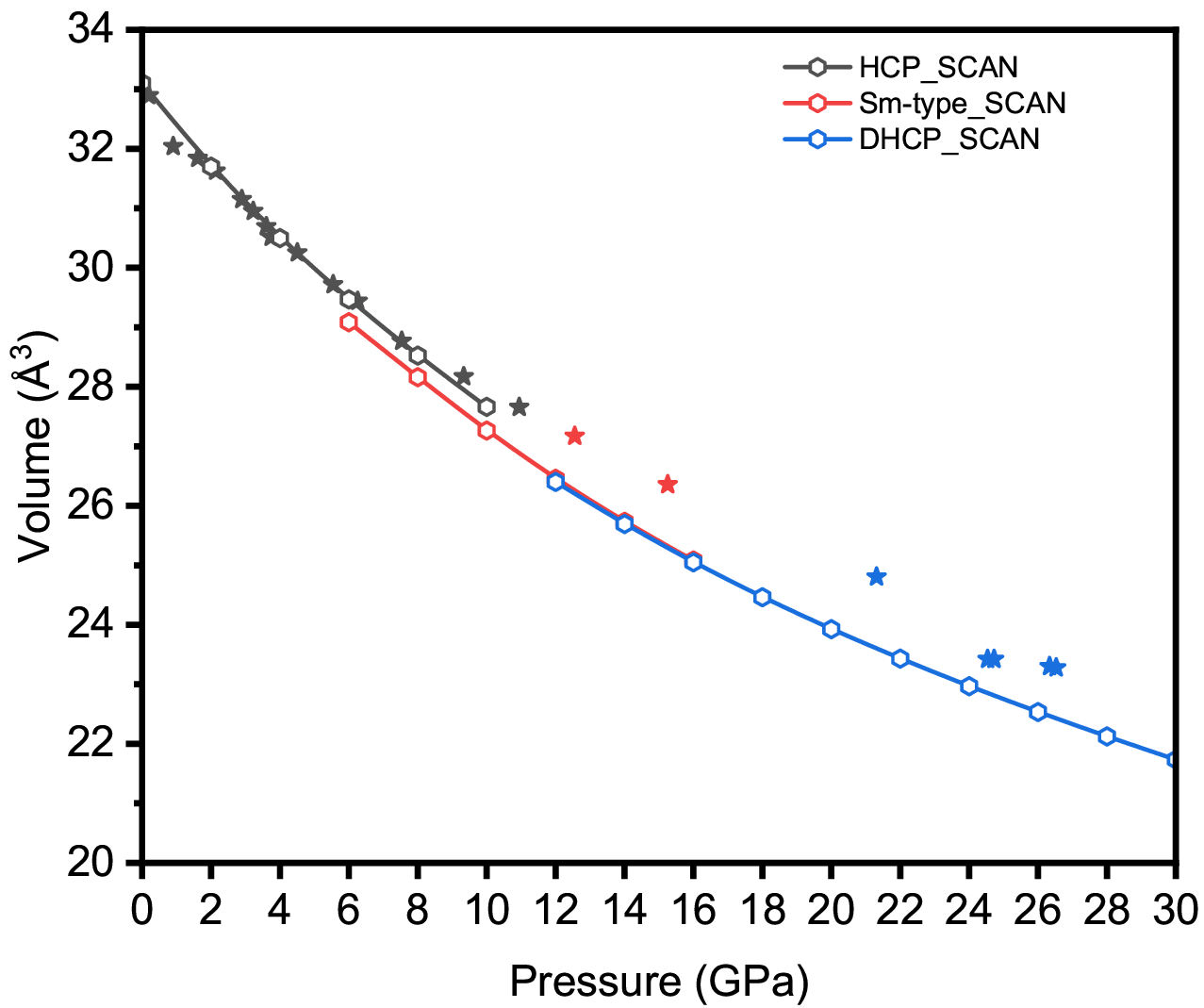}
	\end{minipage}
	\hfill
	\begin{minipage}[t]{0.32\textwidth}
		\raggedright
		{\fontsize{8}{12}\selectfont \textbf{(c)}} \\  
		\includegraphics[width=\linewidth]{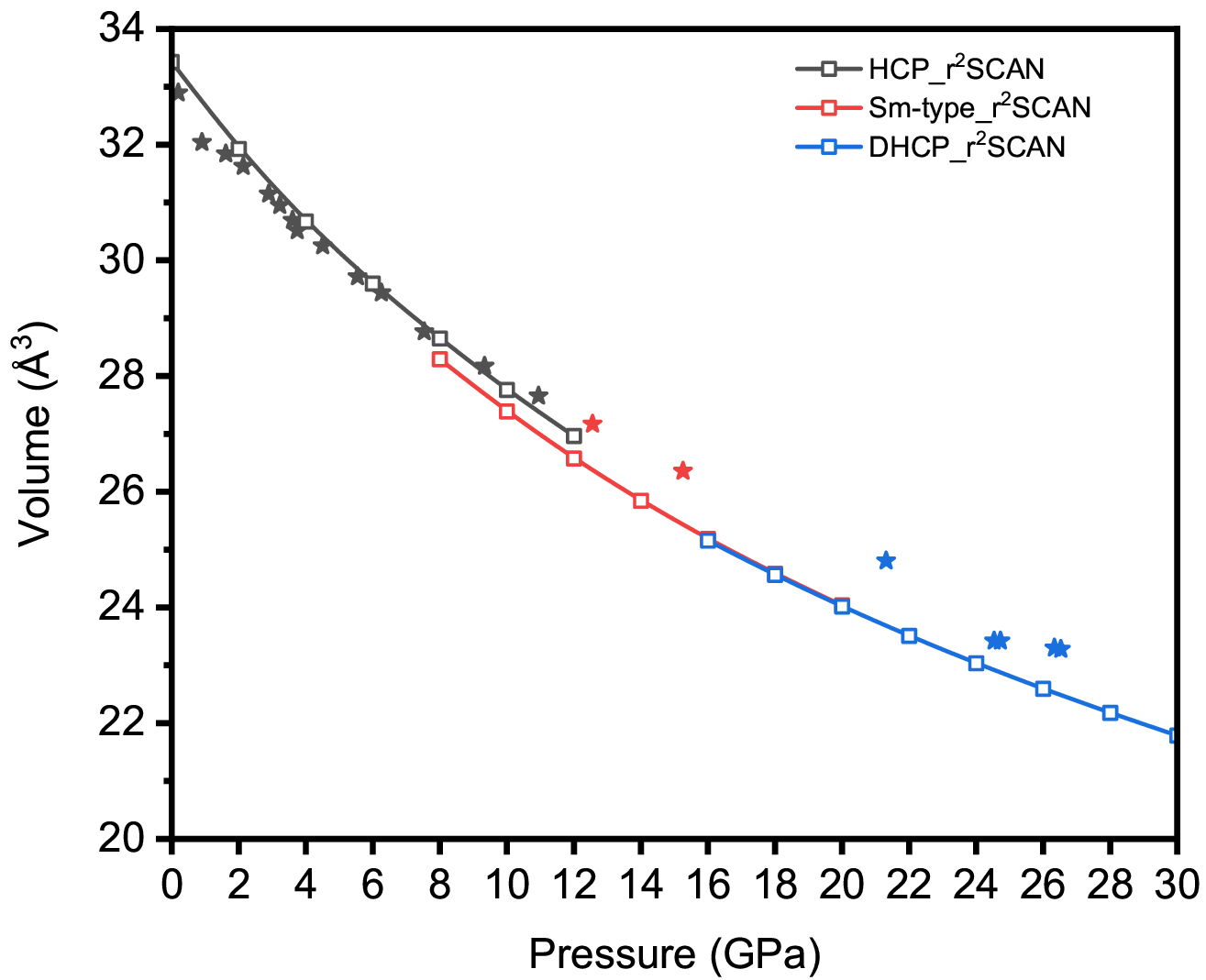}
	\end{minipage}
	
	\vspace{10pt}
	
	\begin{minipage}[t]{0.32\textwidth}
		\raggedright
		{\fontsize{8}{12}\selectfont \textbf{(d)}} \\  
		\includegraphics[width=\linewidth]{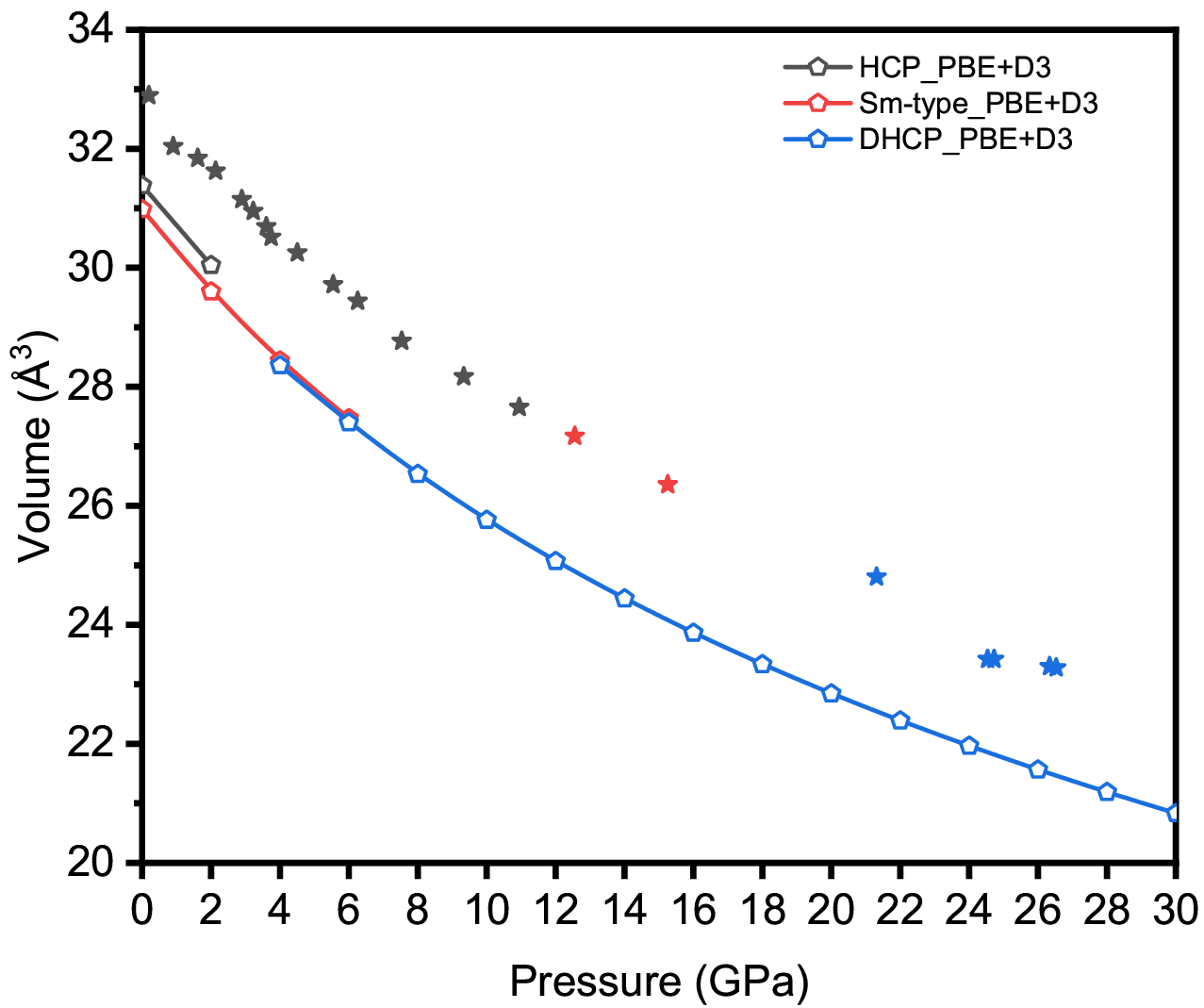}
	\end{minipage}
	\hspace{10pt} 
	\begin{minipage}[t]{0.32\textwidth}
		\raggedright
		{\fontsize{8}{12}\selectfont \textbf{(e)}} \\  
		\includegraphics[width=\linewidth]{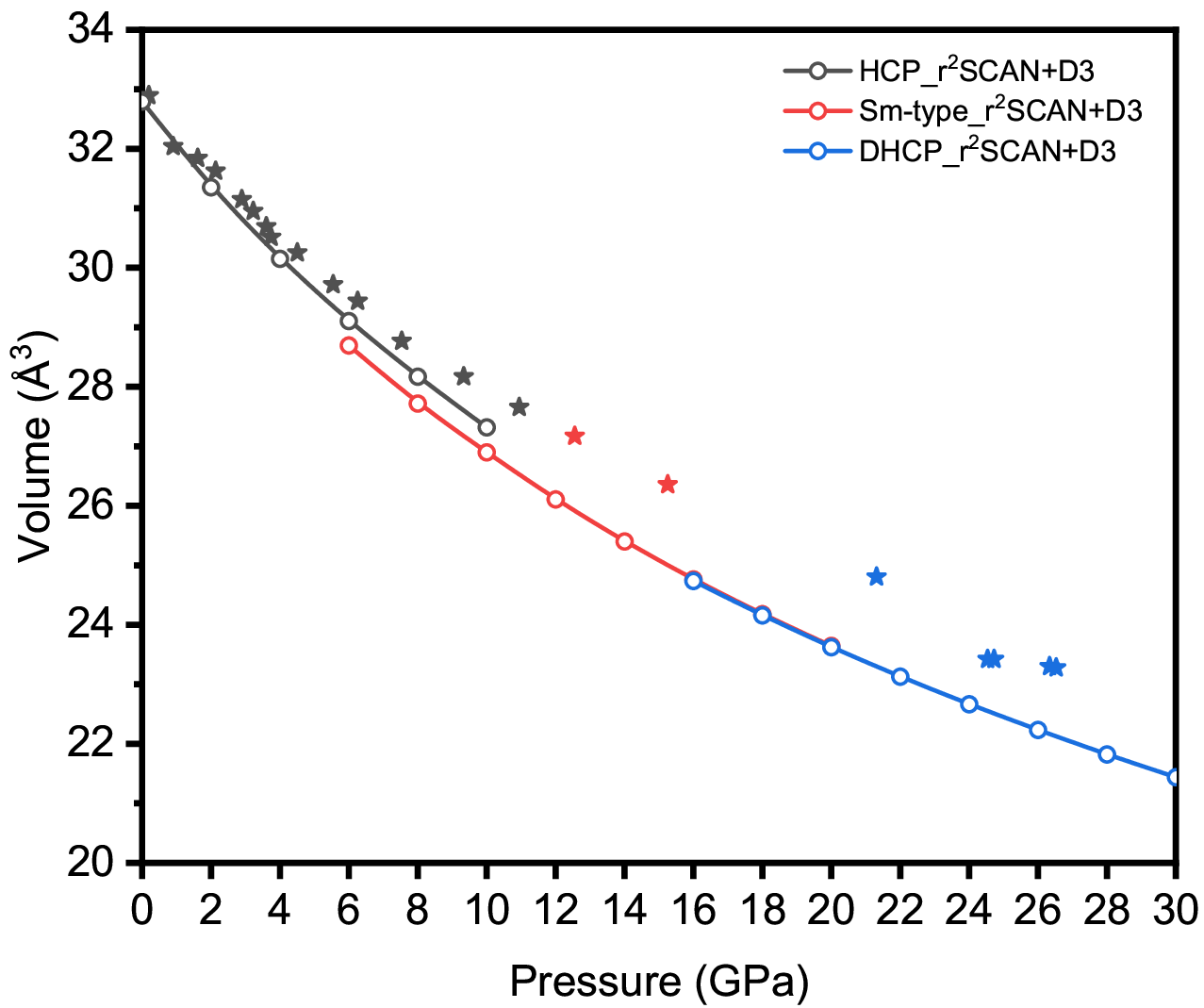}
	\end{minipage}
	
	\caption{The volume-pressure profile using PBE-GGA (a), SCAN-metaGGA (b), r$^2$SCAN-metaGGA (c) and with dispersion corrected D3 level of functionals (d and e).}
	\label{fig:S2}
\end{figure}

\begin{figure}[h]
	\centering
	\includegraphics[scale=0.35]{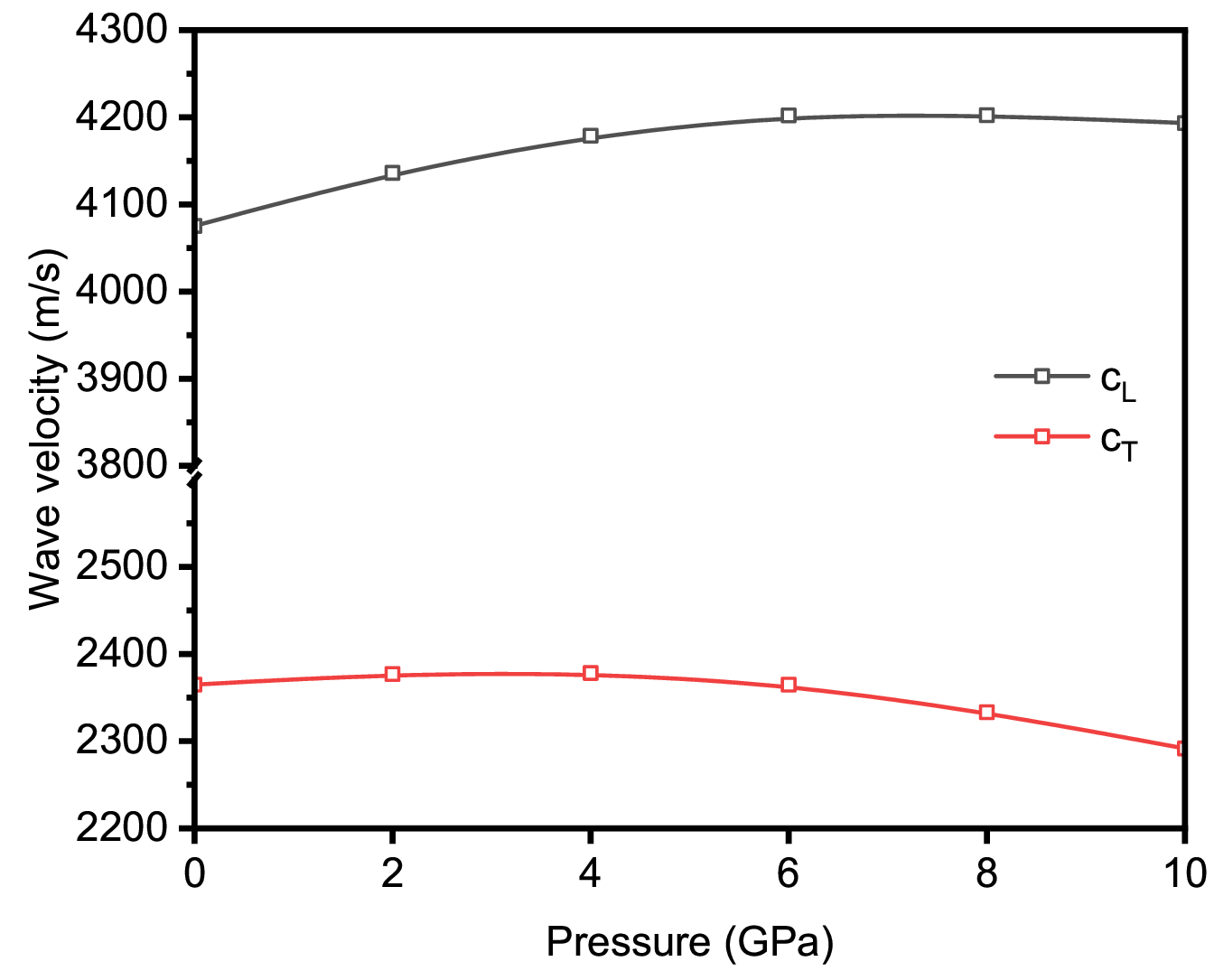}
	\caption{The longitudinal (c$_L$) and transverse (c$_T$) sound wave velocities with respect to the pressure for $hcp$ phase.}
\end{figure}

\begin{figure}[h]
	\centering
	
	\begin{minipage}[t]{0.45\textwidth}
		\raggedright
		{\fontsize{8}{12}\selectfont \textbf{(a)}} \\  
		\includegraphics[width=\linewidth]{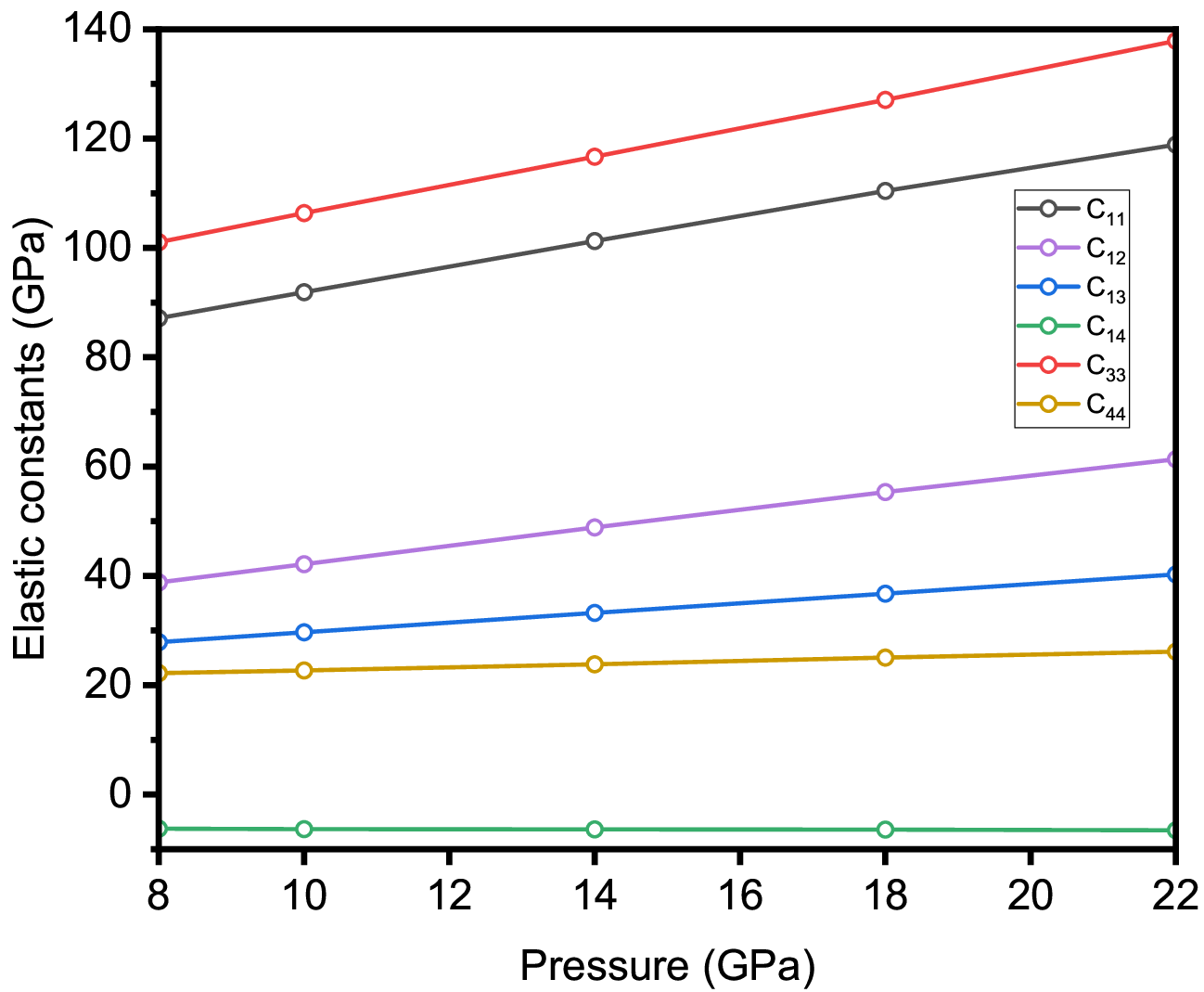}
	\end{minipage}
	\hfill
	\begin{minipage}[t]{0.45\textwidth}
		\raggedright
		{\fontsize{8}{12}\selectfont \textbf{(b)}} \\  
		\includegraphics[width=\linewidth]{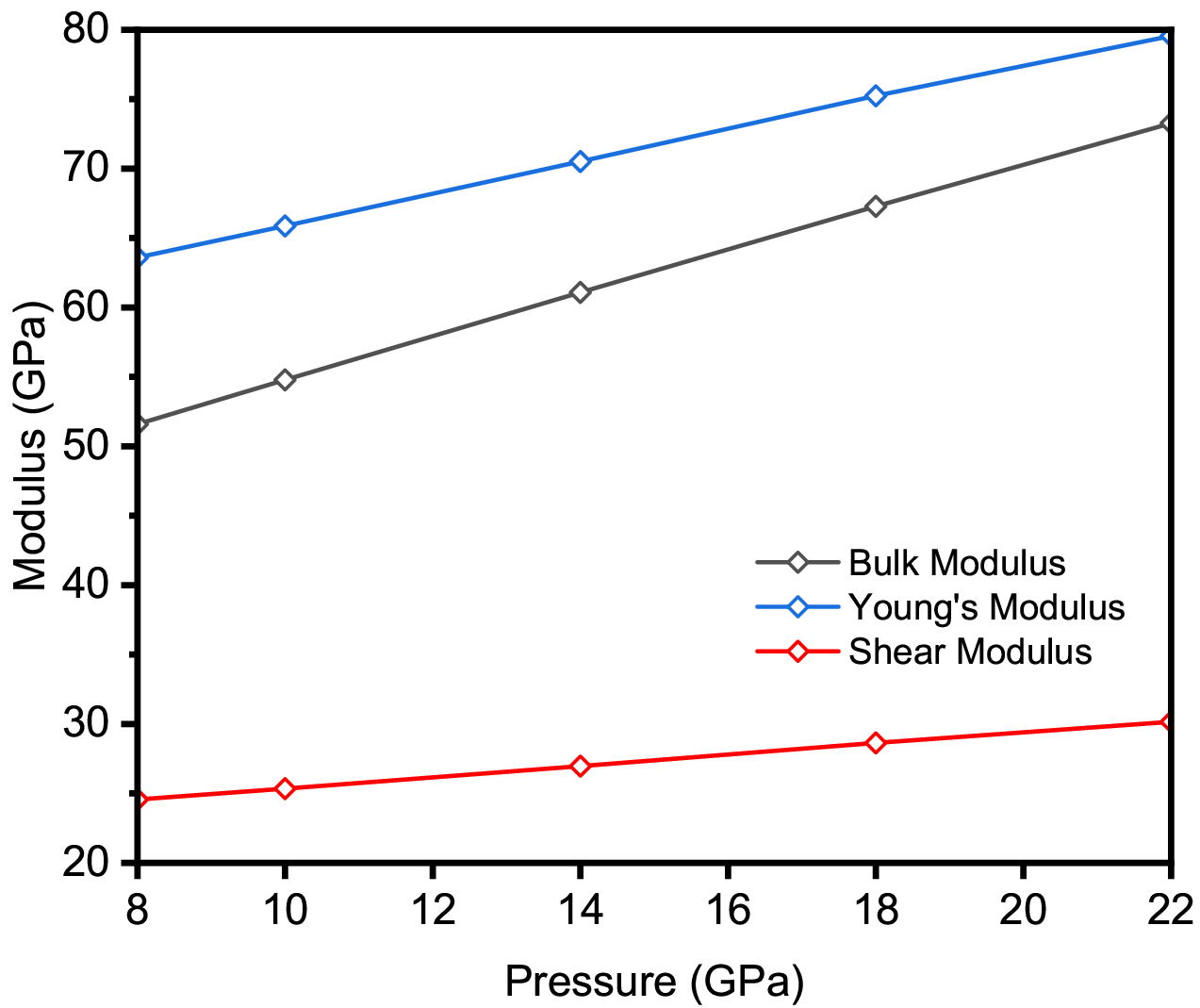}
	\end{minipage}
	\caption{The elastic constants (a) and modulus (b) of $Sm$-$type$ phase over the pressure range.}
	\label{fig:S6}
\end{figure}

\clearpage